\documentclass[sigconf,nonacm]{acmart} 

\AtBeginDocument{%
  }

\setcopyright{acmlicensed} 
\copyrightyear{2018} 
\acmYear{2018} 
\acmDOI{XXXXXXX.XXXXXXX} 
\acmConference[Conference acronym 'XX]{Make sure to enter the correct
  conference title from your rights confirmation email}{June 03--05,
  2018}{Woodstock, NY}  
\acmISBN{978-1-4503-XXXX-X/2018/06}  

\usepackage{tikz}
\usepackage{xurl}
\usepackage[inline]{enumitem}
\usepackage{xspace}
\usepackage{listings}
\usepackage{amsmath}
\usepackage{booktabs}
\usepackage{multirow}
\usepackage{amsfonts}
\usepackage{soul}
\usepackage{subfigure}
\usepackage[table]{colortbl}
\usepackage{xcolor}
\usepackage{versions}
\usepackage{bbm}
\usepackage{pifont}
\usepackage{tcolorbox}
\tcbuselibrary{listings, skins}
\usepackage{soul}

\usepackage[ruled,vlined,noresetcount]{algorithm2e}
\usepackage{caption}
\usepackage[lambda,advantage,operators,probability]{cryptocode}

\newcommand{\aadv}{\ensuremath{\mathcal{A}}\space}
\newcommand{\badv}{\ensuremath{\mathcal{B}}\space}

\definecolor{myOrange}{HTML}{C76000}

\lstdefinelanguage{Tamarin}{
  keywords={rule, let, in, builtins, equations, restriction, lemma,
            all-traces, exists-trace, Fr, In, Out, theory, begin, end, pk, All, Ex, not, h, K, reuse},
  keywordstyle=\color{blue!70}\bfseries,
  keywords=[2]{
    SecretKey, Key, QESignKey, QEVerKey,
    InitTD, InitialInstall,
    TDIsRunning, OperationInputs, CompleteOperation, ReportReady, SignatureReady,
    InputMessage, RunOperation,
    TDProducesReport, MacCheck, QEQuotes, Verified, VrfAccepts,
    PrvMemory, ExpectedState, VrfState, VrfSendsChallenge,
    TDMeasurement, ExpectedSoftware, runOp,
    State, PendingLookup, Trained, train, datasets, SecretOutput
  },
  keywordstyle=[2]\color{myOrange}\bfseries,
  morecomment=[l]{//},
  morecomment=[l]{///},
  morecomment=[s]{/*}{*/},
  columns=fullflexible,
  keepspaces=true,
  sensitive=true
}

\newcounter{tamarinrule}
\newcounter{tamarinlemma}
\renewcommand{\thetamarinrule}{\arabic{tamarinrule}}
\renewcommand{\thetamarinlemma}{\arabic{tamarinlemma}}

\newcommand{\ruleheader}[1]{%
  \refstepcounter{tamarinrule}%
  \label{rule:\thetamarinrule}%
  \textbf{\underline{Rule~\thetamarinrule:} #1}\\%
}

\newcommand{\lemmaheader}[1]{%
  \refstepcounter{tamarinlemma}%
  \label{lemma:\thetamarinlemma}%
  \textbf{\underline{Lemma~\thetamarinlemma:} #1}\\%
}

\lstdefinestyle{tamarinstyle}{
  language=Tamarin,
  basicstyle=\footnotesize\ttfamily,
  frame=single,
  breaklines=true,
  commentstyle=\color{gray},
  escapechar=|,
}

\usepackage{multirow}

\usetikzlibrary{shapes,arrows}
\usetikzlibrary{backgrounds, positioning, fit}
\makeatletter
\tikzset{
    database/.style={
        path picture={
            \draw (0, 1.5*\database@segmentheight) circle [x radius=\database@radius,y radius=\database@aspectratio*\database@radius];
            \draw (-\database@radius, 0.5*\database@segmentheight) arc [start angle=180,end angle=360,x radius=\database@radius, y radius=\database@aspectratio*\database@radius];
            \draw (-\database@radius,-0.5*\database@segmentheight) arc [start angle=180,end angle=360,x radius=\database@radius, y radius=\database@aspectratio*\database@radius];
            \draw (-\database@radius,1.5*\database@segmentheight) -- ++(0,-3*\database@segmentheight) arc [start angle=180,end angle=360,x radius=\database@radius, y radius=\database@aspectratio*\database@radius] -- ++(0,3*\database@segmentheight);
        },
        minimum width=2*\database@radius + \pgflinewidth,
        minimum height=3*\database@segmentheight + 2*\database@aspectratio*\database@radius + \pgflinewidth,
    },
    database segment height/.store in=\database@segmentheight,
    database radius/.store in=\database@radius,
    database aspect ratio/.store in=\database@aspectratio,
    database segment height=0.1cm,
    database radius=0.25cm,
    database aspect ratio=0.35,
}
\makeatother

\usepackage{environ}

\newcounter{definitionctr}

\NewEnviron{mydefinition}[2][]{%
    \refstepcounter{definitionctr}%
    \ifx\\#1\\\else\label{#1}\fi%
    \begin{figure}[h]%
    \vspace{-0.91em}%
    \noindent\fbox{%
        \parbox{0.95\columnwidth}{%
        \small
            \textbf{\underline{Definition~\thedefinitionctr:}} #2%
            \vspace{2pt}\par\noindent%
            \BODY%
        }%
    }%
    \vspace{-0.91em}%
    \end{figure}%
}

\newcommand{\method}{\textsc{PAL*M}\xspace}

\newcommand\change[1]{{\edit{#1}\xspace}}
\renewcommand{\quote}{\texttt{QUOTE}\xspace}
\newcommand{\tdreport}{\texttt{TDREPORT}\xspace}
\newcommand{\modelarch}{\mathcal{M}_{ar}\xspace}
\newcommand{\modelPreTrain}{\mathcal{M}_{tr}\xspace}

\newcommand{\trainconfig}{\mathcal{T}\xspace}
\newcommand{\model}{\mathcal{M}\xspace}

\newcommand{\dtrain}{\mathcal{D}_{tr}\xspace}
\newcommand{\dtest}{\mathcal{D}_{te}\xspace}

\newcommand{\dataset}{\mathcal{D}\xspace}

\newcommand{\out}{\mathcal{O}\xspace}
\newcommand{\inp}{\mathcal{I}\xspace}
\newcommand{\att}{Att\xspace}
\newcommand{\gpuatt}{GPU_{att}\xspace}

\newcommand{\hist}{\ensuremath{\mathcal{H}}}

\newcommand{\intr}{{\ensuremath{\sf{\mathsf{Inr}}}}\xspace}
\newcommand{\tauth}{{\ensuremath{\sf{Tru}}}\xspace}
\newcommand{\prv}{{\ensuremath{\sf{Prv}}}\xspace}
\newcommand{\vrf}{{\ensuremath{\sf{Vrf}}}\xspace}

\newcommand{\chal}{{\textit{Chal}}\xspace}

\newcommand{\adv}{\ensuremath{\sf{\mathcal Adv}}\xspace}
\newcommand{\MSH}{{\texttt{MSH}}\xspace}
\newcommand{\hash}{{\texttt{h}}}
\newcommand{\dHash}{{$\mathtt{h}_{\dataset}$}}
\usepackage{arydshln}

\newcommand{\vruletab}{\vrule width 0.6pt}
\newcommand{\dvruletab}{\vrule width 0.6pt\hspace{1pt}\vrule width 0.6pt}
\newcommand{\baseline}[1]{\colorbox{gray!15}{#1}}
\newcommand{\overhead}[1]{\colorbox{blue!15}{#1}}
\newcommand{\total}[1]{\colorbox{orange!15}{#1}}

\newcommand{\baselinetab}[1]{\cellcolor{gray!15}#1}
\newcommand{\overheadtab}[1]{\cellcolor{blue!15}#1}
\newcommand{\totaltab}[1]{\cellcolor{orange!15}#1}

\newcommand{\attrDistr}{\ensuremath{A_{dist}}}

\usepackage{tikz}
\usepackage{subfigure}
\usepackage{dashbox}
\usepackage{cleveref}

\usetikzlibrary{arrows}
\usetikzlibrary{shapes}
\newcommand*\circled[3]{\tikz[baseline=(char.base)]{
            \node[scale=.7,shape=circle,draw,inner sep=1pt,fill=#2,minimum size=.1cm,text=#3] (char) {#1};}}

\newcommand{\bverb}[2][blue!70]{%
  \textbf{\textcolor{#1}{\lstinline[breaklines=false, basicstyle=\ttfamily]{#2}}}%
}

\newcommand{\overb}[2][orange!90]{%
  \textbf{\textcolor{#1}{\lstinline[breaklines=false, basicstyle=\ttfamily]{#2}}}%
}

\newcommand{\edit}[1]{\textcolor{black}{#1}}

\includeversion{submission}
\excludeversion{arxiv}

\begin{document}

\title{\method: Property Attestation for Large Generative Models}




\author{
{\large{Prach Chantasantitam$^1$, Adam Ilyas Caulfield$^1$, Vasisht Duddu$^1$, Lachlan J. Gunn$^2$, N. Asokan$^{1,3}$}\\
$^1$University of Waterloo, $^2$Aalto University, $^3$KTH Royal Institute of Technology}\\
\normalsize{\texttt{\{pchantas}, \texttt{acaulfield}, \texttt{vasisht.duddu\}@uwaterloo.ca}, \texttt{lachlan@gunn.ee}, \texttt{asokan@acm.org}}
} 


\pagestyle{plain}

\setlist{nolistsep}

\begin{abstract}
\emph{Machine learning property attestations} allow provers (e.g., model providers or owners) to attest properties of their models/datasets to verifiers (e.g., regulators, customers), enabling accountability towards regulations and policies. But, current approaches do not support generative models or large datasets.
%
%
We present \method, a property attestation framework for large generative models, illustrated using large language models. \method defines properties across training and inference, leverages \emph{confidential virtual machines} with \emph{security-aware} GPUs for coverage of CPU-GPU operations, and proposes using \emph{incremental multiset hashing} over memory-mapped datasets to efficiently track their integrity.
We implement \method on Intel TDX+NVIDIA H100 and evaluate it using state-of-the-art models and datasets, showing \method is efficient, incurring <11\% overhead for common operations. Finally, we use the Tamarin Prover symbolic verification tool to formally model \method's property attestation protocol, confirming that its security guarantees are upheld under the defined threat model.

\end{abstract}




\maketitle

\section{Introduction}\label{sec:introduction}

Machine learning (ML) models are widely deployed across domains such as healthcare, finance, and autonomous  systems~\cite{llm_medical,llm_autonomous_vehicle,llm_fin}.
Emerging regulations (e.g., EU's AI Act~\cite{eu_ai_act}) require 
models meet certain \emph{properties}
related to accuracy, training procedures, data provenance, and inference. 
Providing proof of such properties may be challenging for confidential datasets or models. 

\emph{ML property attestation} enables a \emph{prover} \prv (e.g., a model provider) to demonstrate properties about their model or datasets to a \textit{verifier} (\vrf)~\cite{duddu2024attesting}. 
Approaches based on \emph{trusted execution environments} (TEEs)~\cite{sgx,tdx,amd-sev-snp,trustzone}, first proposed in Laminator~\cite{duddu2024laminator}, has been shown to be efficient, scalable, versatile, and secure for classifiers, whereas alternative approaches based on zero-knowledge proofs~\cite{zkpot1,zkpot2,zkcnn,zklora,zkllm,zkmlaas} or re-purposing of inference attacks~\cite{jian_liu_break_PoL,papernot_break_PoL} do not meet these requirements.

But, Laminator does not scale to generative models, such as large language models (LLMs). For example, large datasets cannot reside fully in TEE memory and are \emph{random sampled} from untrusted storage, undermining standard integrity measurement strategies. 

Enabling ML property attestations for generative models requires addressing three challenges:
(1) handling of large datasets outside of TEE memory sampled \emph{randomly},
(2) supporting CPU-GPU computing environments, and
(3) defining property attestation for common operations related to generative models.
We address these challenges with \method, the first framework for \emph{ML property attestations of large generative models}, using LLMs for illustration.

\method employs \textit{incremental multiset hashing}~\cite{msh} to enable compact secure integrity tracking of large, randomly-accessed, datasets, and TEE-aware GPUs to efficiently and securely ensure integrity of heterogeneous operations.
Finally, we define property attestations for a comprehensive set of operations (e.g., fine-tuning, quantization, LLM chat sessions) that do not require revealing confidential datasets or models to \vrf.
We prototype and evaluate \method using Intel TDX and NVIDIA H100.
%
%
%
%
Our contributions are: we
\begin{enumerate}[leftmargin=*]
    \item propose use of incremental multiset hash functions to construct representative measurements of datasets in external storage that are randomly sampled at run-time (Section~\ref{sec:measure-datasets});
    \item define how to measure properties of generative model operations, incorporating GPU attestation evidence without exposing confidential details (Section~\ref{sec:properties});
    \item specify a property attestation protocol that shows how such measurements and outputs can be combined to prove data/models were produced by a \method-equipped CPU-GPU configuration (Section~\ref{sec:attestation}); 
    \item implement a fully functional prototype\footnote{To be released after peer review} of \method built on commercially available CVM and TEE-aware GPU -- Intel TDX and NVIDIA H100 (Section~\ref{sec:setup}); and
    \item show that \method meets desired requirements, including efficiency for common operations, using real-world datasets/models, and security, by formally modeling and verifying \method's property attestation protocol using Tamarin Prover (Section~\ref{sec:evaluation}).
\end{enumerate}

\section{Background}\label{sec:background}


\subsection{Large Language Models}\label{sec:genmodels}

\textbf{Training.}
LLMs use transformer architectures with billions of parameters~\cite{vaswani2017attention,brown2020language}, and are trained to predict the next tokens given previous ones. 
During training, the model learns a probability distribution 
of the next token from a vocabulary to follow a sequence of tokens.
This is often called \emph{pretraining}, as it is typically followed by another phase of training for task-specific optimization/tuning.

\textbf{Tokenizers.} 
Modern tokenizers operate at word, sub-word (e.g., byte pair encoding, WordPiece), or character level to convert natural-language text into numerical representations for use during LLM training and inference.

\textbf{Fine-tuning.}
Since training LLMs is resource- and time-intensive, publicly available trained LLMs are generally fine-tuned on a task-specific dataset. 
Parameter-Efficient Fine-Tuning (PEFT) and Low-Rank Adaptation (LoRA) fine-tune a small subset of model parameters instead of retraining the entire model~\cite{peft, lora}. Alternatively, \emph{adapter layers} are trainable modules that can be attached to a pretrained LLM, enabling task-specific adaptation without altering the core model parameters. 
These approaches are more efficient than fine-tuning the entire LLM.




\textbf{Evaluating LLMs.}
Benchmarks for LLMs cover a variety of tasks and compute relevant metrics.
These tasks span language understanding (e.g., MMLU~\cite{mmlu}), math problems, multiple-choice, classification, and open-ended question answering.
Each task is designed to probe specific capabilities of the model.

\textbf{Chat sessions.} LLMs are commonly used for chatbots, which allow a user to continuously query about a particular task and get the responses from the model that consider prior conversational context (e.g., OpenAI's ChatGPT).
Chat sessions contain multiple user queries and their responses.

\textbf{Memory Management.}
Large generative models require significantly more data than traditional ML models, which often do not fit into main memory (e.g., Pile dataset is over 800 GB~\cite{pile}). Instead of loading the entire data into memory, current LLM libraries (e.g., Hugging Face) support \emph{memory-mapped datasets}~\cite{hf_memory_mapped}, which map the contents of a file into a process’s virtual memory space, allowing it to be accessed like regular memory, rather than through explicit file I/O operations, despite residing on disk. When a dataset is memory-mapped, dataset records are only loaded into memory as needed, significantly reducing memory usage.

\textbf{Data Transformations and Processing:} 
Before a dataset is used, it typically undergoes a series of transformations, including \emph{preprocessing} steps like normalization, concatenation, truncation, tokenization, and splitting into training, validation, and test sets, as well as additional processing for removing corrupted samples and formatting the data to match the expected input structure of the training framework~\cite{hf_preprocess}.
During training, \emph{random sampling} is commonly used to reduce bias introduced by data ordering and to improve model generalization~\cite{robbins1951stochastic}. PyTorch's DataLoader and Sampler abstractions provide randomized sampling~\cite{pytorch_data}. Popular training libraries, such as Hugging Face Transformers, support random sampling and enable it for default training pipelines~\cite{hf_random}.

\subsection{Trusted Execution Environments}\label{sec:tees}

TEEs are a set of architectural components that provide a secure execution environment that is isolated from the ``rich'' host environment. They guarantee that software executing outside of the TEE cannot directly read or write to data inside the TEE's trusted boundary. The exact details of the trusted boundary vary based on the particular approach. For example, Intel 
SGX~\cite{sgx} creates TEE in the user-space \textit{enclaves} and  ARM TrustZone~\cite{trustzone} creates a TEE containing all system resources (e.g., separate trusted OS and applications with dedicated peripherals).

Recent technologies have also introduced \textit{confidential virtual machines} (CVMs), which enable entire VMs to execute within the trust boundary of a TEE (e.g., AMD SEV-SNP~\cite{amd-sev-snp}, Intel TDX~\cite{tdx}, and ARM Realms~\cite{realms}). An advantage provided by CVMs is that they enable executing unmodified and full-feature operating systems without requiring TEE-specific rewrites. Additionally, TEE-aware GPUs have recently emerged (e.g., with NVIDIA H100 GPU). As such, CVM providers have released drivers to support integration with TEE-aware GPUs. This has enabled the potential for heavy computations (e.g., ML training) within a TEE. Both CPUs and GPUs with TEEs typically support \emph{remote attestation} to prove trustworthiness to verifiers.

Intel Trust Domain eXtensions (TDX)~\cite{tdx} enables isolated CVMs in the form of trust domains (TDs). 
Upon creation of a TD, invoked by the untrusted virtual machine manager (VMM), the TDX module computes a measurement of TD components during TD creation, which can be used later for attestation.
To attest TD's state, the TD invokes the TDX module to generate a \texttt{TDREPORT}, which is generated by the CPU computing a MAC over measurements of the TD image, the TDX module image, and any \texttt{REPORTDATA} extended by the TD at runtime. The resulting \texttt{TDREPORT} and MAC are passed by the TDX Module to the quoting enclave (QE), a special-purpose Intel SGX enclave~\cite{intel_tdx_base_spec}. The QE uses contents of \texttt{TDREPORT} to generate a SGX-style \texttt{QUOTE} using a provisioning certification key~\cite{intel_pcs}. Before generating this \texttt{QUOTE}, the QE first validates the MAC, ensuring integrity is maintained when passed by the untrusted VMM. 

\begin{figure}[t]
    \centering
        \scalebox{0.85}{
    \begin{tikzpicture}
        \fill[blue!40!white] (1,0) rectangle (1.25,2);
        \fill[blue!40!white] (3.75,0) rectangle (4,2);
        
        \draw[->, thick] (1.5, 1.5) -- node[above] {\small (1) Challenge} (3.5,1.5);
        \draw[->, thick] (3.5, 0.25) -- node[above] {\small (3) Report State} (1.5, 0.25);
        
        \node[text width=3cm] at (2.35,2.25) {\small \vrf};
        \node[text width=3cm] at (5.15,2.25) {\small \prv};
        \node[text width=3cm] at (5.75,1) {\small (2) Authenticated \\ State Measurement};
        \node[text width=3cm] at (1.0,0.25) {\small (4) Verify};
    \end{tikzpicture}
    }
    \vspace{-0.5em}
    \caption{Traditional remote attestation protocol: a prover (\prv) demonstrates its software integrity to a verifier (\vrf).}
    \vspace{-0.5em}
    \label{fig:RA}
\end{figure}
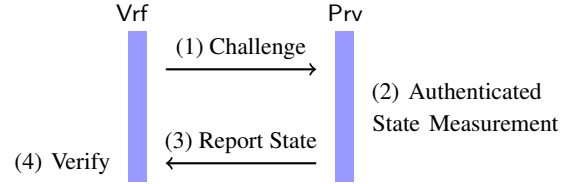

\subsection{Attestation Protocols}\label{sec:ra}

Remote attestation is a protocol in which \vrf challenges \prv to demonstrate that they are installed with the correct configurations~\cite{nunes2024toward}.
The steps for a traditional remote attestation protocol are visualized in Figure~\ref{fig:RA}.
%
The authenticated integrity measurement in step (2) is computed by \prv's \emph{root of trust} (RoT) and is typically a message authenticated code (MAC) or digital signature. Thus, one requirement of remote attestation is that \prv's RoT can securely store and use a private key, implying some level of hardware assistance. 
Hence, TEEs can naturally enable remote attestation with their properties for isolated data and execution. For Intel TDX, step (2) involves computing \texttt{TDREPORT} and \texttt{QUOTE}.

Property attestation~\cite{prop_att} is where \vrf is interested in \textit{properties} of \prv's configuration rather than a specific configuration itself.
Here, \vrf obtains the authenticated integrity measurement from \prv, while also obtaining reference values that correspond to a particular property from a \emph{trusted authority} (\tauth). If the state measurement from \prv is among the reference values obtained from \tauth, and was signed using the expected public key from \prv, \vrf trusts that the property holds on \prv. 
Notably, unlike standard remote attestation, property attestation requires \vrf to obtain reference values from \tauth rather than to recompute them.
Every property of interest is associated with a \emph{measurer} on \prv used to measure (e.g., hash) inputs and outputs corresponding to the property.

The notion of property attestation has been extended recently to cover ML data and processes~\cite{duddu2024attesting,duddu2024laminator,prop_att}. In particular, Laminator~\cite{duddu2024laminator} leverages Intel SGX~\cite{sgx} to attest properties of training data, model, and training process. This works by executing ML operations inside an SGX enclave, measuring the inputs and outputs of the target operation (e.g., training, inference, or dataset statistics), and using SGX’s quote mechanism to return these measurements to the verifier that issued the attestation challenge.
Such attestations can then be used for verifiable ML property cards for transparency (e.g., model card, datasheet, and inference card).

\section{Problem Statement}\label{sec:problem}

Our goal is to design a property attestation framework for generative models
to prove to \vrf that 
a claimed property is (a) correctly measured and (b) not false or manipulated. 

\begin{figure}[t]
    \centering
    \scalebox{0.72}{
    \begin{tikzpicture}
        \fill[blue!40!white] (0.15,0) rectangle (.4,5);
        \fill[blue!40!white] (3.25,0) rectangle (3.5,5);
        \fill[blue!40!white] (6.2,0) rectangle (6.45,5);
        \fill[blue!40!white] (8.8,0) rectangle (9.05,5);

        \node[text width=3cm] at (1.6,5.25) {\small \tauth};
        \node[text width=3cm] at (4.6,5.25) {\small \vrf};
        \node[text width=3cm] at (7.6,5.25) {\small \intr};
        \node[text width=3cm] at (10.2,5.25) {\small \prv};

        \draw[->, thick] (0.55, .85) -- node[above] {\small\parbox{2.5cm}{\centering (6) Send Reference Values}} (3.05, 0.85);

        \node[text width=3cm] at (3.3,0.25) {\small (7) Verify};

        \draw[->, thick] (3.6, 2.5) -- node[above] {\small\parbox{2.5cm}{\centering (4) Request State Measurement}} (6, 2.5);

        \draw[<-, thick] (3.6, 1.5) -- node[above] {\small\parbox{2.5cm}{\centering (5) Report State}} (6, 1.5);
        
        \draw[->, thick] (6.6, 4.5) -- node[above] {\centering \small (1) Request} (8.6,4.5);
        
        \draw[->, thick] (8.6, 3.25) -- node[above] {\centering \small (3) Report State} (6.6, 3.25);    
        
        \node[text width=3cm] at (10.7,4) {\small (2) Authenticated \\ State Measurement};

    \end{tikzpicture}
    }
    \vspace{-2em}
    \caption{Desired property attestation for ML, in which \prv interacts with an Initiator (\intr). \vrf requests evidence from \intr, and may consult a Trusted Authority (\tauth) to obtain reference values of a desired property.}
    \label{fig:prop-att}
\end{figure}
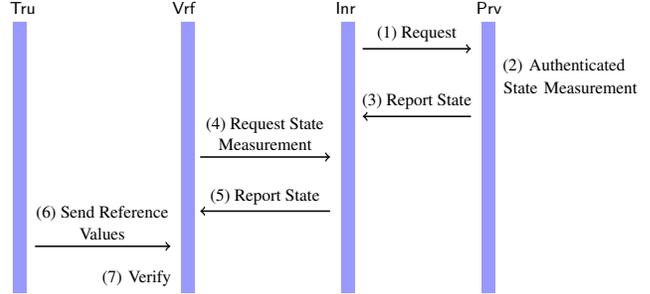

Unlike typical remote attestation, which is an interactive protocol between \vrf and \prv, ML property attestation must be non-interactive with \vrf. 
We propose the protocol depicted in Figure~\ref{fig:prop-att}. \prv interacts with an \textit{initiator} (\intr), who sends a request to \prv for a particular operation, along with other request-specific data (e.g., inputs or a challenge when freshness is required). 
Upon request, \intr makes them available for \vrf, who may consult \tauth for property reference values to verify the evidence. 

\subsection{System Model}\label{sec:sys}
We consider \prv that trains, evaluates, or deploys a model $\model$.
We consider software that is already executing within CVMs, as ML companies today are exploring use of CVMs~\cite{brave2025verifiableprivacy,anthropic2025confidentialinference,openai2024reimaginingsecureinfrastructure},
and major cloud service providers (e.g., Google Cloud and Microsoft Azure) support TEE-aware GPUs~\cite{google_cc, azure_cc}.
We consider Intel TDX and NVIDIA H100 in this work, but our framework can be extended to any CVM or TEE-aware GPU (see Appendix~\ref{sec:discussions}).

Property attestations have the following components:
\begin{itemize}[leftmargin=*]
    \item \textbf{\emph{Operation ($op$)}}: an ML-related computational task over inputs ($\mathcal{I})$ to produce outputs ($\mathcal{O}$).
    \item \textbf{\emph{Property}}: an expression that describes the semantics of some meaningful relationship between $op$, $\mathcal{I}$, and $\mathcal{O}$.
    \item \textbf{\emph{Attestation evidence}}: a verifiable claim that aims to prove whether a property holds true.
\end{itemize}
We assume an attestation mechanism on \prv executes $op$ and produces attestation evidence. 
Obtaining attestation evidence requires (1) measuring $op$, $\mathcal{I}$, $\mathcal{O}$ and (2) reporting those measurements by producing a cryptographic token (e.g., a digital signature) binding them.
For the latter, a signing key is used that must be accessible only to the attestation mechanism.

\subsection{Adversary Model}\label{sec:adv}

We assume an adversary (\adv), in line with the common Intel TDX adversary model~\cite{tdx_security_report}, who has control over the host machine. As a consequence, \adv can access and manipulate any memory or device that is not within the trusted boundaries (e.g., other VMs, disk, I/O peripherals).
The TDX module and TDX-aware CPU are trusted to be implemented correctly and provide their described functionality for inter-TD memory isolation, isolation from host VMM, and hardware-protected measurement and reporting (e.g., via a \texttt{QUOTE}) of TD images.
Since TD creation is managed by \adv-controlled VMM, TDs are not inherently trusted. Hence, TDX supports attestation via \texttt{QUOTE} (see Section~\ref{sec:background}). 

We assume NVIDIA H100 GPU hardware is trusted, which measures and attests to its configuration (e.g., boot ROM, firmware, in confidential computing mode). By default, the interface between a GPU and the TD is not trusted; the GPU and all its memory are brought under the TD’s protection only when the GPU is directly assigned by the VMM~\cite{intel_tdx_base_spec}.
This includes mapping its communication buffers into TD shared memory. TD must obtain an attestation token from the GPU to determine that the assigned device has the anticipated configuration.

At run-time, \adv may attempt to obtain confidential data or interfere with the execution of $op$ in a TD.
\adv also may attempt to forge \texttt{QUOTE}s, either by learning the signing key or by tampering with relevant code/data without detection.
We assume a Dolev-Yao \adv as it relates to the network (i.e., \adv carries the message). Therefore, \adv may attempt to inject or modify messages sent/received by \prv.
We consider denial-of-service attacks out of scope. 

Side channels are prevalent and can be exploited to leak keys~\cite{tdx_sidechannel,wilke2024tdxdown,rauscher2025tdxploit,lee2025t, mishra2025oops,schwarz2017malware}. To mitigate these attacks, additional mechanisms (e.g., constant-time software) can be added; as several works have studied side-channel mitigations~\cite{tdx_module,tdx_rapler,constable2023aex,vanoverloop2025tlblur,blime}, we consider these defenses as orthogonal to the focus of this paper. We also consider physical attacks on hardware (e.g., interposing memory buses~\cite{tee-fail} and or physically swapping GPUs) as out of scope.

\begin{figure*}[t!]
    \centering
    \includegraphics[width=0.75\textwidth]{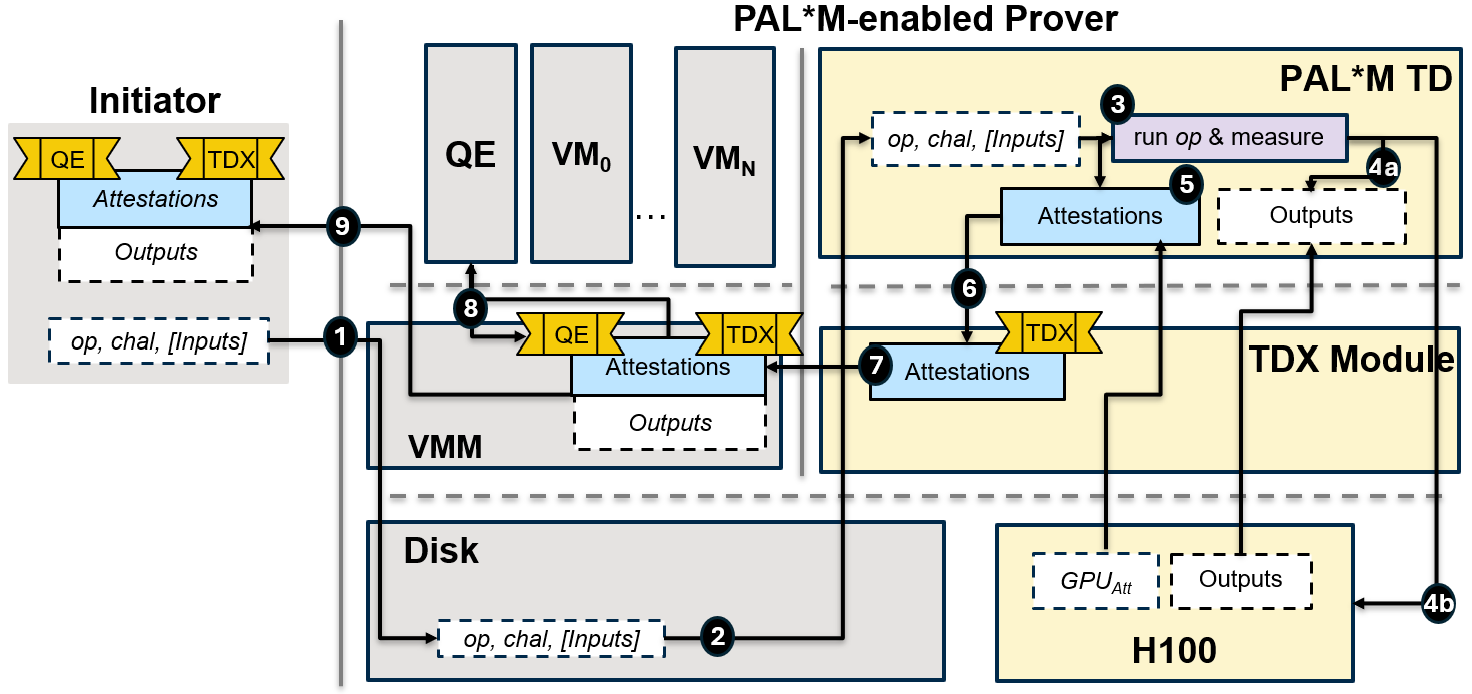}
    \vspace{-0.5em}
    \caption{High-level overview of \method-enabled property attestation. \intr interacts with untrusted components of \prv, which saves inputs to disk. \method reads and measures inputs, runs the operations (including GPU), and measures all CPU and GPU outputs. All measurements are extended to \texttt{REPORTDATA}, which is used to as input to obtain \texttt{TDREPORT} from TDX Module. Finally, a \texttt{QUOTE} is generated by invoking the Quoting Enclave (QE) and returned to \intr.}
    \label{fig:overview}
\end{figure*}

\subsection{Challenges and Requirements}

Prior works~\cite{duddu2024laminator,exclavefl,atlas,slapp,attestable_audits} demonstrate methods to attest ML models using TEEs, but require operations to stay within the one TEE.
For example, Laminator~\cite{duddu2024laminator} requires that all computations stay inside an SGX enclave to ensure integrity. 
However, this is not possible with reasonable performance of LLM operations. For example, large datasets may require datasets to reside in external storage during operation and require accelerators (e.g., GPUs or NPUs) for reasonable performance (recall Section~\ref{sec:background}).
Although it is possible to outsource computations from a TEE to a standard GPU~\cite{slalom,jang2019heterogeneous}, this adds significant overhead and would not preserve integrity of property measurements without additional measures.
Furthermore, no prior work defines \textit{properties} of generative models (e.g., from operations like fine-tuning, quantization, and chat sessions). Therefore, it is unclear how to perform property attestation that is relevant for \vrf while protecting \prv's confidential assets.

Due to these challenges, prior work cannot apply to properties of generative models, requiring another approach to address the following concerns: 
\begin{enumerate}[leftmargin=*]
\item accounting for random sampling of large datasets from (untrusted) external storage devices;
\item handling operations that depend on both CPU and GPU with reasonable computational overhead;
\item defining how to measure properties of generative models in a way that is relevant for \vrf and does not reveal \prv's confidential assets.
\end{enumerate}

Based on these challenges, we consider the following requirements for designing \method:
\begin{enumerate}[label={\textbf{[R\arabic*]}},leftmargin=*]
    \item\label{efficient}\emph{Efficient}: low-overhead compared to confidential computing in the same environment;
    \item \label{robust}\emph{Secure}: a malicious \prv cannot generate false attestations.
    \item\label{scalable}\emph{Scalable}: supports large numbers of provers/verifiers;
    \item\label{versatile}\emph{Versatile}: supports a variety of properties; requires minimal effort for custom property measurement; can be ported to any CVM/TEE-based GPU configuration.
\end{enumerate}
\begin{figure*}[t!]
    \centering
    \includegraphics[width=0.7\textwidth]{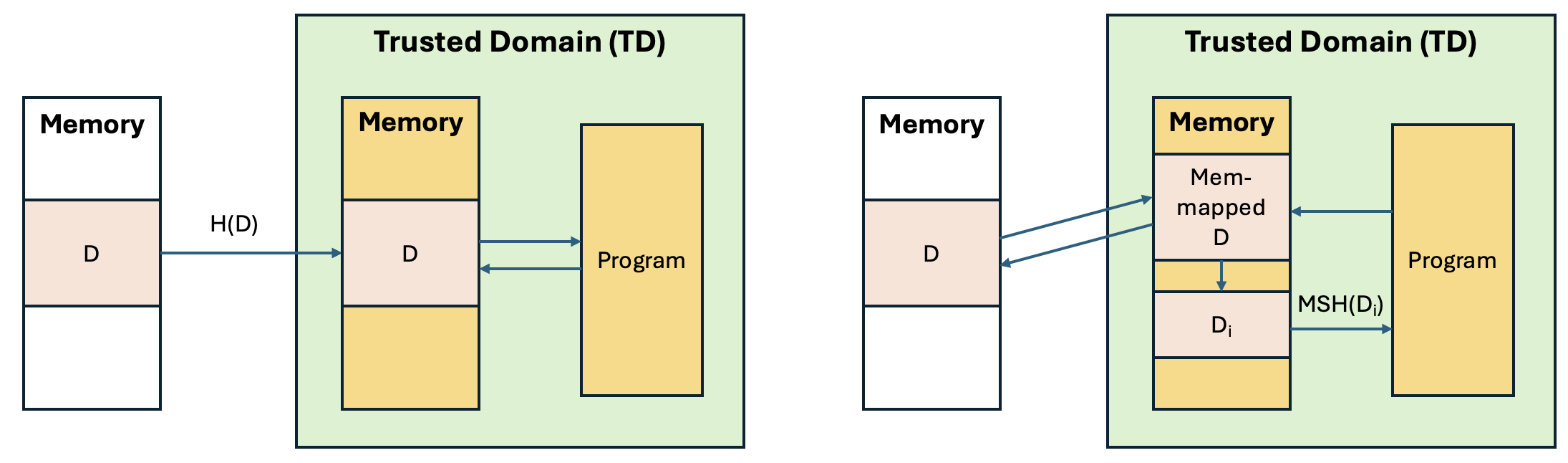}
    \vspace{-0.5em}
    \caption{Depiction of \method’s handling of \textit{in-memory} (left) and \textit{memory-mapped} (right) dataset. In the in-memory approach, the entire dataset is loaded and measured once at load time. In the memory-mapped approach, individual records are measured when accessed using a multiset hash to ensure consistent measurements regardless of sampling order.} 
    \label{fig:mem-type}
\end{figure*}

\section{\method Design}\label{sec:design}

\subsection{Overview of \method}\label{sec:overview}

\method enables \prv to attest properties of generative models through (1) design of a TD that effectively measures the operation corresponding to a particular property and (2) use of attestation support from Intel TDX and NVIDIA H100 to construct authenticated evidence of these measurements. 
Figure~\ref{fig:overview} shows a high-level overview of \prv's workflow with \method and the interaction with \intr. 

In \circled{1}{black}{white}, \intr requests a property attestation by specifying: (a) the requested operation ($op$), (b) an attestation challenge ($\chal$), and any inputs that should be used (e.g., datasets, models, configuration parameters, inference query).
Upon receipt, \prv's VMM saves all data to disk.
In \circled{2}{black}{white}, inputs are fetched by \method into TD memory according to the dataset access strategy (discussed further in Section~\ref{sec:measure-datasets}). Based on $op$, the corresponding function will start to execute in \circled{3}{black}{white}. While executing, outputs will be obtained by a combination of CPU execution \circled{4a}{black}{white} and GPU execution \circled{4b}{black}{white}. Operation outputs and attestation measurements are obtained \circled{5}{black}{white}. When a GPU is used, the attestation measurements include a GPU-attestation token ($GPU_{att}$).

After outputs are obtained, \method TD leverages TDX mechanisms to convert property attestation measurements into authenticated evidence. 
As a first step, it invokes the TDX module in \circled{6}{black}{white} to generate an authenticated \texttt{TDREPORT} containing \method attestation measurements. To complete the quote process, \texttt{TDREPORT} is passed to the VMM in \circled{7}{black}{white}, which invokes the QE in \circled{8}{black}{white} to convert \texttt{TDREPORT} into a \texttt{QUOTE}. Finally, \texttt{QUOTE} is returned to \intr in \circled{9}{black}{white}. \intr will provide \texttt{QUOTE} and outputs to requesting \vrf, who consults \tauth (e.g., Intel and NVIDIA) to perform evidence verification. Protocol details are described further in Section~\ref{sec:attestation}.

By design, \method generates \texttt{QUOTE}s that attest to properties of generative model operations, while addressing key challenges:
\begin{enumerate}[leftmargin=*]
    \item \method leverages \emph{incremental multiset hash functions}~\cite{msh} to ensure large datasets are accurately reflected in attestation evidence despite residing in untrusted external storage devices and being randomly sampled;
    \item \method ensures measurements of both CPU and GPU configurations are captured in attestation evidence for properties that depend on CPU and GPU.
\end{enumerate}

Building upon these capabilities, we show how to define, measure, and attest to properties of the following generative model operations (from Section~\ref{sec:genmodels}):
\begin{enumerate}[leftmargin=*]
    \item dataset transformations and distributional statistics;
    \item  model pretraining, optimization (e.g., fine-tuning, quantization), and evaluation;
    \item model interactions (e.g., inference or chat sessions).
\end{enumerate}

\subsection{Dealing with Large Datasets}\label{sec:measure-datasets}

Given the large size of LLM training datasets, 
ML software frameworks use two common approaches for efficient dataset handling: (1) loading the entire dataset \textit{in-memory} or (2) accessing the dataset outside of TD protected memory with \textit{memory mapping}. 
Although in-memory has an advantage for integrity measurements, frameworks such as Hugging Face adopt and default to using memory-mapped datasets to minimize the memory footprint of large datasets that cannot fit entirely into memory~\cite{hf_memory_mapped}. 

For accurate attestation, \method must measure the dataset (or its items) upon use, since \adv could tamper with data in external storage at any time (recall Section~\ref{sec:adv}).
Figure~\ref{fig:mem-type} shows how \method handles this for \textit{in-memory} and \textit{memory-mapped} datasets.


\textbf{Case 1: In-memory Dataset}.
For integrity measurement, this is the ideal approach when sufficient TD memory is available. Since the TD memory is protected, data within cannot be tampered. As such, the dataset can be measured once during loading, as long as it stays in memory for the entire operation. As shown in  Figure~\ref{fig:mem-type} (left), our framework loads the entire dataset into memory, measures it, and uses it with the operation thereafter.
Since one hash operation is required, the overhead is minimal (discussed further in Section~\ref{sec:evaluation}). 

\textbf{Case 2: Memory-mapped Dataset}.
In contrast to the in-memory case, the dataset resides on the disk, and a virtual address mapping is established in memory (see Figure~\ref{fig:mem-type}: right). As records are sampled, the OS reads just the requested items from disk into memory on demand. Since the dataset stays on disk, it is vulnerable to tampering at run-time, resulting in inconsistencies between what is measured and what is used (i.e., TOCTOU attacks~\cite{bratus2008toctou}).

To account for this, we measure each record at the time it is sampled, accumulating a measurement of the dataset in use. Since training algorithms often \emph{randomly sample} records, a straightforward approach (e.g., a hash chain over accessed records) would be unsuitable, as different sampling orders would result in different measurements.  
Based on this insight, we employ an \textit{incremental multiset hash} (\MSH)~\cite{msh}.

\MSH is a cryptographic primitive that produces a fixed-size hash over an unordered collection. It is suitable for settings in which the order of elements is irrelevant, but integrity and compactness are required. By accumulating an \MSH over the sampled records, \method accurately measures the dataset that is used at run-time.
Additionally, this approach has low memory overhead since one multiset hash value is maintained. 

However, hashing records alone is insufficient if \adv can reorder them at run-time. Such reordering may affect model characteristics (e.g., fairness~\cite{ganesh2023impact}) while the resulting \MSH would be unchanged, since it is insensitive to sampling order. Hence, instead of accumulating a record ($d_i$) directly, we accumulate the concatenation of each record with its index ($d_i | i$). This approach makes it possible to detect any reordering by \adv at run time, as the resulting \MSH will differ due to an index mismatch with the requested index.

We use the \texttt{Mset-Mu-Hash} multiplicative multiset hash construction (from Theorem 1 in~\cite{msh}).
Let $B$ denote a set of bit vectors of length $m$, $M$ a multiset of elements of $B$, and $M_{b}$ be the number of times $b \in B$ is in the multiset $M$. 
Let $q$ be a large prime power and consider computations in the field $GF(q)$. Then, let $\hash$ be a poly-random function performing $\hash: B \rightarrow GF(q)$
The multiplicative multiset hash $\MSH(M)$ is defined as: $\MSH(M) = \prod_{b \in B} \hash(b)^{M_b}$

Compared to straightforward dataset hashing, this approach incurs additional overhead due to iterative hash and modular operations based on the hardness of discrete logarithm problem, requiring a large prime. However, it ensures no other countermeasures are required to account for random sampling, as 
any permutation of the full set of dataset read operations will result in the same final $\MSH$.
We note that we chose to use the \texttt{Mset-Mu-Hash} construction due to its simplicity and widely available library support, but any \MSH construction could be used.
For example, \MSH construction based on elliptic curves can be used for more efficiency gains, but have other limitations (further details in Appendix~\ref{apdx:msh-comparison}). 
Even with \texttt{Mset-Mu-Hash}, overhead of \MSH is negligible compared to computation-heavy ML tasks (shown in Section~\ref{sec:evaluation}).

It is possible that expected dataset measurement from \tauth is not \texttt{MSH}. Consequently, when memory-mapped datasets are in use, \method must also enable methods for \vrf to trace back to original dataset measurements.

\subsection{Property Measurements for Large Generative Models}\label{sec:properties}


Recall from Section~\ref{sec:sys} that obtaining property attestation evidence requires (1) obtaining measurements of $\{op, \mathcal{I}, \mathcal{O}\}$ and (2) reporting those measurements by producing an authentication token. In this section, we discuss the former. 

We categorize properties measured by \method as one of the following based on $op$: dataset property, model property, model interaction property. Based on this, we focus on the following components of  property measurement (subset of those from Section~\ref{sec:sys}):
\begin{itemize}[leftmargin=*]
    \item \textbf{\textit{Input ($\inp$)}}: inputs (e.g., datasets, models, queries, or other parameters) used by $op$.
    \item \textbf{\textit{Output ($\out$)}}: outputs (e.g., prompt response, model, or dataset) produced by $op$.
    \item \textbf{\textit{Property}}: an expression that describes semantics (either natural language or mathematical expression) of the relationship between $\{op, \mathcal{I}, \mathcal{O}\}$.
    \item \textbf{\textit{Attestation Measurement ($\att$)}}: Measurements that reflect the property under the adversary model (from Section~\ref{sec:adv}).
\end{itemize}
A summary of frequent notation in this section is in Table~\ref{tab:notation}.

\begin{table}[t]
    \caption{Notation Summary}
    \vspace{-1.2em}
    \label{tab:notation}
    \centering
    \renewcommand{\arraystretch}{0.9}
    \footnotesize
    \begin{tabular}{c p{0.6\columnwidth}}
        \bottomrule
    
        \toprule
        \textbf{Symbol} & \textbf{Definition} \\
        \bottomrule
    
        \toprule
        $\inp$, $\out$, $\att$ & Set of inputs, outputs, and attestation measurements \\
        $P_{dist}$ & Distributional property \\
        $\MSH$ & Multiset hash function \\
        $\hash$ & Standard cryptographic hash function (e.g., SHA3) \\
        \dHash & Dataset hash (either $\MSH$ or $\hash$ based on memory management strategy from Sec.~\ref{sec:measure-datasets}) \\
        $\gpuatt$ & Attestation token from TEE GPU \\
        $\dataset$ & Dataset \\
        $\dataset_{opt}$, $\dtest$ & optimization dataset, test dataset \\
        $\dtrain$ & Training dataset \\
        $\dataset_{pre}$ & Preprocessed dataset \\
        $\modelarch$ & Model architecture \\
        $\trainconfig$ & Training configuration \\
        $\model$ & Model \\
        $\model_{tok}$ & Tokenizer \\
        $id_{opt}$ & Optimization type \\
        $\model_{adp}$ & Adapter model \\
        $\dataset_{opt}$ & Optimization dataset \\
        $[\cdot]^*$ & Optional components \\
        $metric$ & Evaluation metric \\
        $q_i,r_i$ & The $i$-th inference query and response \\
        $\mathcal{H}$ & Session history (i.e., set of all query-response pairs [($q_0,r_0$), ... , $(q_{i-1},r_{i-1})$]) that are used as additional context to produce $q_{i}$ for $r_{i}$ \\
        \bottomrule
        
        \toprule
    \end{tabular}
    \vspace{-1.3em}
\end{table}

\subsubsection{GPU Attestation.} For all operations that require GPU, $\att$ must include $\gpuatt$ to reflect the complete environment (i.e., GPU driver, firmware/VBIOS, and other configurations) in which the computations were performed. 
Therefore, $\gpuatt$ is included as an optional input (denoted with $[\cdot]^*$) for all operations described in this section.
Beyond obtaining $\gpuatt$, a secure channel must be established prior to any operation. This is achieved through SPDM-based session key exchange during the TD instantiation~\cite{spdm,gu2025nvidia}, which cryptographically binds all subsequent run-time interactions to the attested GPU instance. The session establishment is also included in the CVM's attestation (i.e., the \texttt{QUOTE}). Together, $\gpuatt$ and SPDM session ensure that the GPU in use is a legitimate confidential GPU and that it remains so throughout the operation.


\subsubsection{Dataset Properties}
We discuss dataset properties based on the following operations: dataset preprocessing, calculation of dataset attribute distribution, and dataset measurement binding.

Dataset preprocessing (from Section~\ref{sec:background}) may occur prior to training or inference.
We define attestation measurement for preprocessing in Definition~\ref{att:preproc}. For preprocessing, $\mathcal{I}$ contains an input dataset ($\dataset$) and $\mathcal{O}$ contains an output preprocessed dataset ($\dataset_{pre}$).
%
A property measurement of preprocessing asserts that $\dataset_{pre}$ was the result of computing a preprocessing algorithm ($\operatorname{PreProc}$) on $\dataset$.
Finally, $\att$ contains measurements of $\dataset$ and $\dataset_{pre}$ using the dataset hash function (\dHash): either a standard cryptographic hash function ($\hash$) or a \MSH, based on whether $\dataset$ and $\dataset_{pre}$ are in-memory or memory mapped, respectively. 


\begin{mydefinition}[att:preproc]{Preprocessing}
    \textbf{Input ($\inp$):} $\dataset$\\
    \textbf{Output ($\out$):} $\dataset_{pre}$\\
    \textbf{Property:} $\dataset_{pre} = \operatorname{PreProc(\dataset)}$ \\
    \textbf{Att. Measurement ($\att$):} \dHash($\dataset$), \dHash($\dataset_{pre}$), $[\gpuatt]^*$
\end{mydefinition}

Dataset attribute distribution calculations~\cite{duddu2024attesting} reflect a distribution of an attribute across a dataset (e.g., word lengths/frequency for LLMs).
We define attestation measurement for calculation of attribute distribution in Definition~\ref{att:dist}.
For this operation, $\mathcal{I}$ contains $\dataset$, and $\mathcal{O}$ contains an attribute distribution (\attrDistr). 
$\att$, containing $\hash(\attrDistr)$ and \dHash$(\dataset)$, asserts the property that \attrDistr was obtained by an algorithm ($\operatorname{Dist}$) that computes $\dataset$'s attribute distribution.



\begin{mydefinition}[att:dist]{Attribute Distribution}
\textbf{Input ($\inp$):} $\dataset$\\
\textbf{Output ($\out$):} \attrDistr\\
\textbf{Property:} $\attrDistr = \operatorname{Dist}(\dataset$) \\
\textbf{Att. Measurement ($\att$):} \hash(\attrDistr), \dHash($\dataset$), $[\gpuatt]^*$
\end{mydefinition}

It is possible that reference values offered by \tauth are produced by $\hash$ while an operation requires memory-mapping, resulting in a measurement based on $\MSH$. 
To account for this scenario, \method supports measurement binding defined in Definition~\ref{att:mapping}. 


\begin{mydefinition}[att:mapping]{Measurement Binding}
\textbf{Input ($\inp$):} $\dataset$,\\
\textbf{Output ($\out$):}  \hash($\dataset$), \texttt{MSH}($\dataset$) \\
\textbf{Property:}  \hash($\dataset$) and \texttt{MSH}($\dataset$) were produced from $\dataset$\\
\textbf{Att. Measurement ($\att$):}  $\hash(\hash(\dataset) || \MSH(\dataset))$, $[\gpuatt]^*$
\end{mydefinition}

For measurement binding, $\mathcal{I}$ contains $\dataset$, and $\mathcal{O}$ contains the measurements of $\dataset$ that should be bound: $\hash(\dataset)$ and $\MSH(\dataset)$.
A property that $\hash(\dataset)$ and $\MSH(\dataset)$ were produced from the same dataset is asserted through $\att$ containing $\hash(\hash(\dataset) || \MSH(\dataset))$.



\subsubsection{Model Properties}
We present model properties related to training and evaluation. 
Among model training operations, we consider model training itself and post-training model optimizations.

\textbf{Training Properties.}
%
First, we consider training by building on prior work~\cite{duddu2024laminator}.
We define attestation measurements for training in Definition~\ref{att:proofPreTrain}.
For training, $\mathcal{I}$ includes the specified model architecture ($\modelarch$), training dataset ($\dtrain$), training configuration ($\trainconfig$), which specifies hyperparameters, and a tokenizer ($\model_{tok}$). Then, $\mathcal{O}$ contains the trained model  ($\modelPreTrain$). 
$\att$ contains measurements of ($\trainconfig$, $\modelarch$, $\modelPreTrain$, $\model_{tok}$) using $\hash$ and $\dataset$ using \dHash, asserting
a property describing that $\modelPreTrain$ was obtained by a training algorithm ($\operatorname{Train}$) taking $\dtrain$, $\trainconfig$, $\modelarch$, $\model_{tok}$.


\begin{mydefinition}[att:proofPreTrain]{Training}
\noindent\textbf{Input ($\inp$)}: $\modelarch$, $\dtrain$, $\trainconfig$, $\model_{tok}$\\
\noindent\textbf{Output ($\out$)}: $\modelPreTrain$ \\
\textbf{Property:}  $\modelPreTrain = \operatorname{Train}(\dtrain, \trainconfig, \modelarch)$ \\
\textbf{Att. Measurement ($\att$)}: $\hash(\modelPreTrain)$, 
$\hash(\modelarch)$, 
\dHash$(\dtrain)$, 
$\hash(\trainconfig)$, $\hash(\model_{tok})$, $[\gpuatt]^*$
\end{mydefinition}

Definition~\ref{att:proofOpt} defines attestation measurements for post-training model optimizations (e.g., fine-tuning, quantization, pruning).
For model optimization, $\mathcal{I}$ contains the following required inputs: $\model$, $\trainconfig$, $\model_{tok}$,  and the optimization type ($id_{opt}$). 
Optional components of $\mathcal{I}$ include a base adapter layer ($\model_{adp}$) and additional datasets ($\dataset_{opt}$).
For this operation, $\mathcal{O}$ contains the optimized model $(\model_{opt})$.
$\att$ contains all components of $\mathcal{I}$ and $\mathcal{O}$ using $\hash$ and \dHash, reflecting its assertion that:
\begin{itemize}[leftmargin=*]
    \item $\operatorname{Opt}$ is selected from a set of known optimization functions ($\mathcal{F}_{opt})$ based on $id_{opt}$;
    \item $\model_{opt}$ is the result of $\operatorname{Opt}$ taking all components of $\mathcal{I}$.
\end{itemize}


\begin{mydefinition}[att:proofOpt]{Weight Optimizations}
\noindent\textbf{Input ($\inp$)}: $\model$, $\model_{tok}$, $\trainconfig$, $id_{opt}$, $[\model_{adp}$, $\dataset_{opt}]^*$ \\
\noindent\textbf{Outputs ($\out$)}: 
$\model_{opt}$ \\
\textbf{Property:} 
$\operatorname{Opt} = \mathcal{F}_{opts}[id_{opt}]$ \\
$\model_{opt} = \operatorname{Opt}(\model, \model_{tok}, \trainconfig, [\model_{adp}$, $\dataset_{opt}]^*)$ \\
\textbf{Att. Measurement ($\att$):} $\hash(\model_{opt})$, $\hash(\model)$, $\hash(\model_{tok})$, $\hash(\trainconfig$), $\hash(id_{opt})$, $[\gpuatt, \hash(\model_{adp})$, \dHash$(\dataset_{opt})]^*$
\end{mydefinition}


\textbf{Evaluation Properties.}
As described in Section~\ref{sec:background}, generative models have various benchmarks to produce metrics reflecting a model's effectiveness according to a particular criteria.
Attestation measurement for model evaluation is defined in Definition~\ref{att:accatt}.
For model evaluation, $\mathcal{I}$ includes $\model$, $\model_{tok}$, and a test dataset ($\dtest$), while $\mathcal{O}$ contains the evaluation \textit{metric}.
%
Measurements of $\model$ and $\model_{tok}$ using $\hash$  and $\dtest$ using \dHash\xspace in $\att$ assert a property about the evaluation: that \textit{metric} was produced by an evaluation operation ($\operatorname{Eval}$) taking $\model$, $\model_{tok}$, and $\dtest$. 


\begin{mydefinition}[att:accatt]{Evaluation}
\noindent\textbf{Input ($\inp$)}: $\model$, $\model_{tok}$, $\dtest$\\
\noindent\textbf{Output ($\out$)}: $metric$ \\
\textbf{Property:} $metric \leftarrow \operatorname{Eval}(\model, \model_{tok}, \dtest)$ \\
\textbf{Att. Measurement ($\att$)}: $\hash(\model), \hash(\model_{tok}),$ \dHash$(\dtest)$, \\
$\hash(metric)$, $[\gpuatt]^*$ 
\end{mydefinition}


\subsubsection{Inference-time Properties}

We consider two cases of inference-time properties which includes interactions between client and model: one prompt-response inference and a \textit{chat session} of multiple prompt-response inference pairs. 

Attestation measurements for a single inference are defined in Definition~\ref{att:inference}. An inference query ($q$) is one component of $\mathcal{I}$ along with $\model$ and $\model_{tok}$, while the $\mathcal{O}$ contains the inference response ($r$). 
Measurements of $q$, $\model$, $\model_{tok}$, and $r$ using $\hash$ assert the property that $q$ was obtained by passing $r$ first through $\model_{tok}$ and then $\model$.


\begin{mydefinition}[att:inference]{Single Inference}
\textbf{Input ($\inp$):} $q$, $\model$, $\model_{tok}$\\
\textbf{Output ($\out$):} $r$ \\
\textbf{Property:} $r = \model(\model_{tok}(q))$ \\
\textbf{Att. Measurement ($\att$):} \hash($q$), \hash($\model$),  \hash($\model_{tok}$), \hash($r$), $[\gpuatt]^*$
\end{mydefinition}


Attestation measurements reflecting the influence of a chat session on a query-response (i.e., a \emph{session inference}) are defined in Definition~\ref{att:session}. 
For a session inference operation, $\mathcal{I}$ contains the same data as the single inference operation, while $\mathcal{O}$ contains $r$ and the chat session history ($\mathcal{H}$): a sequential list of all $(q_i,r_i)$ pairs of the current chat session, including $(q,r)$.
Measurements of $r$, $\mathcal{H}$, $q$, $\model_{tok}$, and $\model$  using $\hash$ assert the property that $r$ was obtained from passing $q$ prepended with $\mathcal{H}$ without $(q,r)$ through $\model_{tok}$, followed by passing that result through $\model$.

\begin{mydefinition}[att:session]{Session Inference}
\textbf{Input ($\inp$):} $q$, $\model_{tok}$, $\model$\\
\textbf{Output ($\out$):}
$r$, $\hist$ \\
\textbf{Property:} $r = \model(\model_{tok}(\hist \setminus [q,r]~||~q))$\\
\textbf{Att. Measurement ($\att$):} $\hash(q)$, $\hash(\model_{tok})$, $\hash(\model)$, $\hash(r)$, $\hash(\hist)$, \\ $[\gpuatt]^*$
\end{mydefinition}



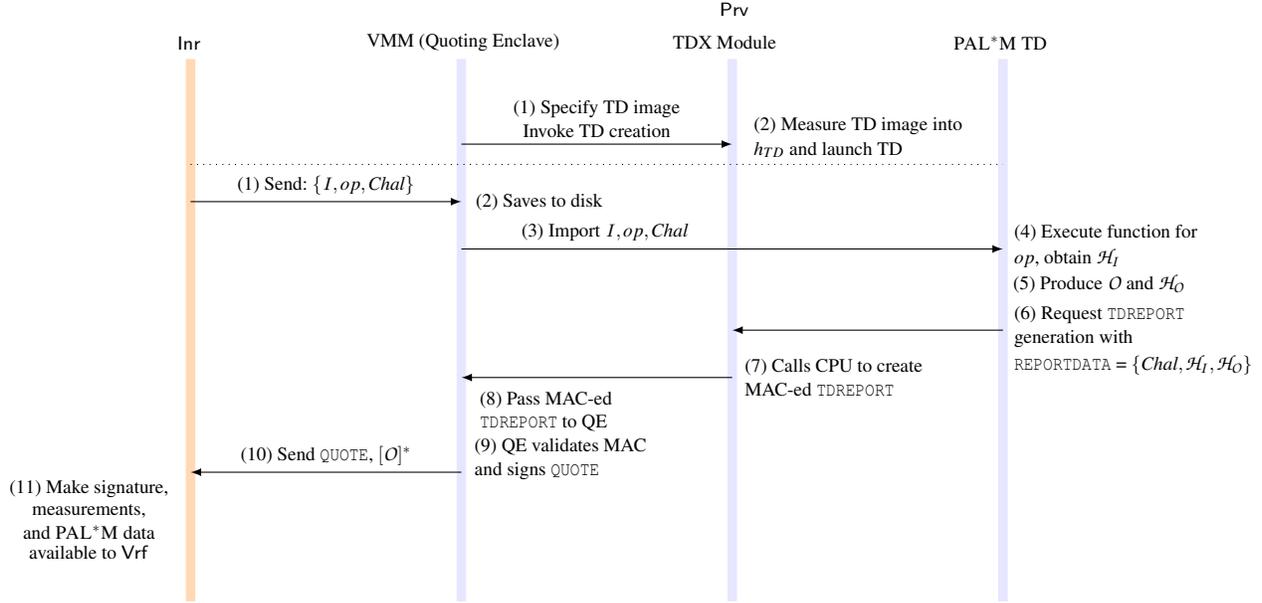
\begin{figure*}[t]
\centering
\footnotesize
\begin{tikzpicture}[>=latex,scale=0.9, every node/.style={scale=0.9}]
\def\ytop{0}   
\def\ybottom{-7} 

\def\xvrf{0}
\def\xvmm{4}
\def\xtdx{8}
\def\xtd{12}

\def\y{0}
\def\dy{-0.7}


\filldraw[fill=orange!30!white, draw=orange!30!white] (\xvrf-.0625,\ytop) rectangle (\xvrf+.0625,\ybottom);
\filldraw[fill=blue!10!white, draw=blue!10!white] (\xtdx-.0625,\ytop) rectangle (\xtdx+.0625,\ybottom);
\filldraw[fill=blue!10!white, draw=blue!10!white] (\xvmm-.0625,\ytop) rectangle (\xvmm+.0625,\ybottom);
\filldraw[fill=blue!10!white, draw=blue!10!white] (\xtd-.0625,\ytop) rectangle (\xtd+.0625,\ybottom);

\node[left] at (\xvrf+.25,\ytop+.25) {\intr};
\node[left] at (\xtdx+.35,\ytop+.75) {\prv};
\node[left] at (\xvmm+1.55,\ytop+.25) {VMM (Quoting Enclave)};
\node[left] at (\xtdx+.75,\ytop+.25) {TDX Module};
\node[left] at (\xtd+.75,\ytop+.25) {\method TD};

\draw[->] (\xvmm,\y-1.25) -- (\xtdx,\y-1.25) node[midway, above] {
\shortstack{(1) Specify TD image \\Invoke TD creation}};
 \node[left] at (\xtdx+3.5,\y-1.15) {\shortstack[l]{(2) Measure TD image into \\$h_{TD}$ and launch TD}};
\pgfmathsetmacro\y{\y+\dy}

\pgfmathsetmacro\y{\y+\dy} 

\draw[dotted] (\xvrf,\y-.15) -- (\xtd,\y-.15);

\pgfmathsetmacro\y{\y+\dy}
\draw[->] (\xvrf,\y) -- (\xvmm,\y) node[midway, above] {(1) Send: $\{\mathcal{I}, op, \chal\}$};
\node[align=left] at (\xvmm+1.15,\y) {(2) Saves to disk};

\pgfmathsetmacro\y{\y+\dy}
\draw[->] (\xvmm,\y) -- (\xtd,\y); 
\node[left] at (\xvmm+3.45,\y+.25) {(3) Import $\mathcal{I}, op, \chal$};

\pgfmathsetmacro\y{\y+\dy+.75}
\node[align=left] at (\xtd+1.52,\y) {\shortstack[l]{(4) Execute function for \\$op$, obtain $\mathcal{H_I}$}};

\pgfmathsetmacro\y{\y+\dy+.15}
\node[align=left] at (\xtd+1.5,\y) {\shortstack[l]{(5) Produce $\mathcal{O}$ and $\mathcal{H_O}$}};

\pgfmathsetmacro\y{\y+\dy}
\node[align=left] at (\xtd+2.125,\y-.15) {\shortstack[l]{(6) Request \texttt{TDREPORT} \\ generation with \\\texttt{REPORTDATA} = $\{op, \chal, \mathcal{H_I}, \mathcal{H_O}\}$}};
\draw[->] (\xtd,\y) -- (\xtdx,\y);

\pgfmathsetmacro\y{\y+\dy}
\node[align=left] at (\xtdx+1.5,\y) {
\shortstack[l]{(7) Calls CPU to create\\ MAC-ed \texttt{TDREPORT}}};
\draw[->] (\xtdx,\y) -- (\xvmm,\y);

\node[align=left] at (\xvmm+1.25,\y-.5){\shortstack[l]{(8) Pass MAC-ed \\ \texttt{TDREPORT} to QE}};
\pgfmathsetmacro\y{\y+\dy}
\node[align=left] at (\xvmm+1.5,\y-.5) {
\shortstack[l]{(9) QE validates MAC \\and signs \texttt{QUOTE}}
};

\pgfmathsetmacro\y{\y+\dy}
\draw[->] (\xvmm,\y) -- (\xvrf,\y) node[midway, above] {\shortstack{(10) Send \texttt{QUOTE}, $[\mathcal{O}]^*$}};

\pgfmathsetmacro\y{\y+\dy}
\node[align=left] at (\xvrf-1.5,\y) {\shortstack{(11) Make signature, \\measurements, \\and \method data  \\ available to \vrf}};

\end{tikzpicture}
\caption{\method Property Attestation Protocol between \colorbox{orange!20!white}{\intr} and \colorbox{blue!10!white}{\prv}. \intr sends request to \prv containing untrusted VMM, Quoting Enclave, trusted TDX Module, and attested \method TD. \prv responds after performing the requested operation, measuring relevant inputs and ouputs, and generating both a  \texttt{TDREPORT} and \texttt{QUOTE}. }
\label{fig:protocol}
\end{figure*}

\subsection{From Measurement to Attestation}\label{sec:attestation}

For each $op$, the corresponding components of $Att$, as defined in Section~\ref{sec:properties}, are used to construct property attestation evidence. We refer to each property attestation as \emph{proof of <$op$>}. A table of operations and resulting property attestations is shown in Appendix~\ref{apdx:property-table}.
In this section, we describe how \method interacts with other system components and leverages Intel TDX features to respond to \intr requests with property attestation evidence. 

Figure~\ref{fig:protocol} shows the property attestation protocol enabled by \method. Here, we describe the interactions between \intr and \prv containing a VMM (including QE), TDX module, and \method-based TD. We note there are several possible interactions between \intr and \vrf (e.g., \intr publishes attestations to a \textit{transparency log}~\cite{exclavefl,atlas}). We discuss implications of alternative strategies in Appendix~\ref{sec:discussions}.

In an offline phase, \prv initializes and measures the \method TD. In step (1), VMM specifies the TD image that should be used. In step (2), the Intel TDX architecture and TDX module measure the TD image into $h_{TD}$ and launch it. 

After initialization, the online protocol can begin. In step (1), \intr sends a property attestation request to \prv containing $op$, an attestation challenge ($\chal$), and a set of input assets ($\mathcal{I}$). 
For large input assets (e.g., datasets or models), \intr's request may specify repositories from which \prv should download them. Or, \intr's request may specify that \prv use a dataset/model owned by \prv. 
Regardless of these specificities, \prv saves all data from the request to disk in step (2) upon receipt. In step (3), the \method TD imports request data and all data required for $op$ into its memory. Datasets are imported depending on whether \prv is using a memory-mapped or an in-memory dataset (from Section~\ref{sec:measure-datasets}).

\method TD reads $op$ to execute in step (4). During execution, it obtains a set of input measurements ($\mathcal{H_{I}}$). After completing the operation, \method will obtain a set of outputs ($\mathcal{O}$) and corresponding measurements ($\mathcal{H_O}$) in step (5). 
$\mathcal{H_{I}}$ and $\mathcal{H_{O}}$ correspond to components of $Att$ for each $op$ described in Section~\ref{sec:properties}.

Next, \method prepares the request for \texttt{TDREPORT} generation in step (6). To do so, it assigns \texttt{REPORTDATA} input used in \texttt{TDREPORT} generation (recall Section~\ref{sec:tees}) as the concatenation of the following data: $\{op, \chal, \mathcal{H_I}, \mathcal{H_O}\}$. Upon receiving \texttt{REPORTDATA}, the TDX module will create a \texttt{TDREPORT} in step (7) by invoking the CPU. During this step, the CPU will compute a MAC over $h_{TD}$, measurements of the TDX module, and all data extended into \texttt{REPORTDATA} by \method. In step (8), the MAC-ed \texttt{TDREPORT} is passed to VMM, who passes it to QE in step (9).
%
Upon receipt, QE:
\begin{itemize}[leftmargin=*,nosep=*]
    \item verifies the MAC on \texttt{TDREPORT} is valid, to ensure VMM did not tamper with \texttt{TDREPORT} it received from the TDX Module;
    \item uses \texttt{TDREPORT} to produce the \texttt{QUOTE} containing all measurements from \texttt{TDREPORT} plus a signature ($sig$).
\end{itemize}
In step (10), \prv sends \texttt{QUOTE} and optionally $\mathcal{O}$ (if it is not confidential) to \intr, who sends them to \vrf in step (11). 

Upon receiving \texttt{QUOTE} and (optionally $\mathcal{O}$) from \intr, \vrf parses the response to perform verification. First, it checks \textit{sig} to determine whether \texttt{QUOTE} came from \prv. Next, it checks $\chal$ to determine the freshness of the response. The type of $\chal$ that is effective depends on the operation (discussed further in Appendix~\ref{sec:discussions}).

Assuming both checks pass, \vrf continues to inspect the measurements of the TD image, TDX module image, and $\gpuatt$ (by obtaining reference values from Intel and NVIDIA).
Assuming this passes, \vrf continues to inspecting $\mathcal{H_{I}}$, $\mathcal{H_{O}}$, and $\mathcal{O}$.
If $\mathcal{O}$ was sent to \vrf, it checks whether $\mathcal{H_{O}}$ corresponds to $\mathcal{O}$. 
Finally, \vrf inspects $\mathcal{H_{I}}$ and $\mathcal{H_{O}}$, consulting reference values from \tauth as needed, to determine if the desired property related to $op$ is upheld. 
 
Like prior work~\cite{duddu2024laminator}, property attestations introduced in this work can be used to construct chains of attestations, enabling verifiers to derive composite guarantees spanning the full ML life-cycle. For example, combining a proof of fine-tuning (attesting that  $\model_{opt}$ was fine-tuned on $\dataset_{opt}$ that was preprocessed with $PreProc(\dataset_{opt})$ algorithm) with proof of inference (attesting that $\model_{opt}$ produces $\out$ with an inference query $q$).
\section{Experiment Setup}\label{sec:setup}

\subsection{Datasets, Models, and Metrics}\label{sec:datasets}\label{sec:models}\label{sec:metrics}

We use the following datasets in our experiments:
\begin{itemize}[leftmargin=*,nosep=*]
    \item \overb{BookCorpus}~\cite{bookcorpus}: a collection of 74 million unpublished novels. We use the full \overb{BookCorpus} dataset for proof of training and attribute distribution, while for proof of preprocessing and proof of binding, we use a 1/100 subset to reduce experiment running time. For training, we preprocess the full \overb{BookCorpus} dataset by tokenizing and concatenating text into sequences of 1024 tokens.
    \item \overb{yahma/alpaca-cleaned}~\cite{alpaca}: a instruction-following dataset containing over 52,000 instruction-response pairs
    aimed at improving model alignment.
    We use this dataset for proof of fine-tuning to provide a dataset variety other than \overb{BookCorpus}.
    \item \overb{MMLU}~\cite{mmlu}: a multiple-choice question-and-answer (Q\&A) dataset covering 57 fields. It is widely used to assess general-purpose intelligence in language models. We use all 57 fields for our proof of evaluation experiments.
    \item \overb{WMT14}\overb{(DE - EN)}~\cite{wmt14}: a test set of WMT14 German-English to evaluate LLM translation capability. This dataset is used to calculate the BLEU score for our proof of evaluation experiment.
    \item \overb{CoQA}~\cite{reddy2019coqa}: a conversational Q\&A dataset containing stories with context-dependent Q\&A. For proof of inference, we select the first question of the first ten records. For the proof of session inference, we use all ten records and select the first five questions from each, simulating a session with 50 prompts.
\end{itemize}




For models, we use \bverb{GPT-2}~\cite{radford2019language} (124M parameters; 0.46 GB) for proof of training. For fine-tuning, inference, and evaluation, we use three popular instruction-tuned LLMs:
\begin{itemize}[nosep=*]
    \item \bverb{Llama-3.1-8B}~\cite{llama3} (8B parameters, 15 GB),
    \item \bverb{Gemma-3-4B}~\cite{gemma3} (4B parameters, 8 GB), and 
    \item \bverb{Phi-4-Mini}~\cite{phi4mini} (3.8B parameters; 7.15 GB).
\end{itemize}
We choose \bverb{Llama-3.1-8B} for its 8 billion parameters, while both \bverb{Gemma-3-4B} and \bverb{Phi-4-Mini} have $\approx$4 billion parameters, allowing us to show how the overhead of \method varies across different model sizes and architectures. To demonstrate a realistic fine-tuning workflow, we fine-tune all three models at full precision using LoRA~\cite{lora}. For quantization, we focus on \bverb{Llama-3.1-8B}. 


\begin{figure*}[t]
\hspace{-0.03\textwidth}
\begin{minipage}{0.60\textwidth}
\raggedleft

\captionof{table}{Performance for proofs of dataset attribute distribution and preprocessing using \overb{BookCorpus} (Full) and \overb{BookCorpus} (1/100).}
    \vspace{-1em}
    \centering
    \footnotesize
    
    \renewcommand{\arraystretch}{1.35}
    \setlength{\tabcolsep}{7pt}
    \scalebox{0.85}{
    \begin{tabular}{l!{\vruletab}c!{\vruletab}c!{\dvruletab}c!{\vruletab}c}
    \bottomrule
    
    \toprule
    \textbf{Proof of} &
    \multicolumn{2}{c!{\dvruletab}}{\textbf{Attribute Distribution}} &
    \multicolumn{2}{c}{\textbf{Preprocessing}} \\
    \bottomrule
    
    \toprule
    \textbf{Dataset} &
    \multicolumn{2}{c!{\dvruletab}}{\textbf{\overb{BookCorpus} (Full)}} &
    \multicolumn{2}{c}{\textbf{\overb{BookCorpus} (1/100)}} \\
    
    \textbf{Memory Variant} &
    \textbf{In-memory} & \textbf{Memory-mapped} &
    \textbf{In-memory} & \textbf{Memory-mapped} \\
    \bottomrule
    
    \toprule
    \textbf{Total Time} &
    \totaltab{339.16 min} & \totaltab{576.97 min} &
    \totaltab{237.70 s} & \totaltab{378.62 s} \\
    \midrule
    
    \textbf{Baseline} &
    \baselinetab{339.09 min} & \baselinetab{331.73 min} &
    \baselinetab{236.66 s} & \baselinetab{236.12 s} \\
    \textbf{Mem Usage (GB)} &
    \baselinetab{5.47} & \baselinetab{4.33} &
    \baselinetab{0.12} & \baselinetab{0.04} \\
    \midrule
    
    \textbf{Raw Meas.} &
    \overheadtab{3.06 s} & \overheadtab{392.68 min} &
    \overheadtab{0.04 s} & \overheadtab{236.06 s} \\
    \textbf{Attestation} &
    \overheadtab{1.06 s} & \overheadtab{0.80 s} &
    \overheadtab{0.90 s} & \overheadtab{0.79 s} \\
    \midrule

    \textbf{Raw Meas. Overhead (\%)} &
    \overheadtab{0.015} & \overheadtab{68.06} & 
    \overheadtab{0.06} & \overheadtab{62.35} \\ 
    \textbf{Att. Overhead (\%)} &
    \overheadtab{0.0052} & \overheadtab{0.0023} &
    \overheadtab{0.38} & \overheadtab{0.21} \\
    \textbf{Total Observed Overhead (\%)} &
    \overheadtab{0.021} & \overheadtab{42.50} & 
    \overheadtab{0.44} & \overheadtab{37.64} \\ 
    \bottomrule
    
    \toprule
    \end{tabular}
    }

    \label{tab:dataset_attestation}

\end{minipage}
\hspace{0.005\textwidth}
\begin{minipage}{0.35\textwidth}
\raggedright
    \captionof{table}{Performance for proof of training with \bverb{GPT-2} on preprocessed \overb{BookCorpus}.}
    \vspace{-1em}
    \centering
    \renewcommand{\arraystretch}{1.25}
    \resizebox{1.1\columnwidth}{!}{%
    \begin{tabular}{l!{\vruletab}c!{\vruletab}c}
    \bottomrule
    
    \toprule
    \textbf{Dataset} & \multicolumn{2}{c}{\textbf{\overb{BookCorpus} (Full)}}  \\
    \textbf{Memory Variant} & \textbf{In-memory} & \textbf{Memory-mapped} \\
    \bottomrule

    \toprule
    \textbf{Total Time} & \totaltab{401.65 min} & \totaltab{425.03 min} \\
    
    \midrule
    \textbf{Baseline} & \baselinetab{401.60 min} & \baselinetab{401.20 min} \\
    \textbf{Mem Usage (GB)} & \baselinetab{2.74} & \baselinetab{0.04} \\
    
    \midrule
    \textbf{Raw Meas.} & \overheadtab{2.27 s} & \overheadtab{23.86 min} \\
    \textbf{Attestation} & \overheadtab{1.11 s} & \overheadtab{0.96 s} \\
    
    \midrule
    \textbf{Raw Meas. Overhead (\%)} & \overheadtab{0.009} & 
    \overheadtab {5.47} \\
    \textbf{Att. Overhead (\%)} & \overheadtab{0.005} & \overheadtab {0.006} \\
    \textbf{Total Observed Overhead (\%)} & \overheadtab{0.014} & 
    \overheadtab {5.61} \\
    \bottomrule
    
    \toprule
    \end{tabular}
    }
    \label{tab:pretrain}
        \vspace{-1em}

\end{minipage}
\end{figure*}


We use the following metrics to capture both the overall overhead from \method and the main sources of overhead:
\begin{itemize}[leftmargin=*,nosep=*]
\item \textbf{Total Time:} the total execution time spent on the task.
\item \textbf{Baseline:} the execution time on a TDX-enabled machine without measurement from \method. We omit a non-TDX baseline, as prior work has evaluated the overhead of TDX itself~\cite{confidential_llm_tdx_h100}.
\item \textbf{Memory Usage:} total memory usage by the dataset.
\item \textbf{Raw Measurement Time:} \edit{time spent purely hashing inputs and outputs (with $\hash$ or \MSH).}
\item \textbf{Attestation:} The time spent by Intel TDX and NVIDIA H100 to produce \texttt{QUOTE} and $\gpuatt$, respectively.
\item \textbf{Raw Measurement Overhead (\%):} \edit{The total raw measurement overhead from \method as a percentage of the total time.}
\item \textbf{Attestation Overhead (\%):} The total attestation overhead from \method as a percentage of the total time.
\item \textbf{Total Observed Overhead (\%):} \edit{The overall observed overhead as a percentage when comparing the total execution time of \method to the baseline.}
\end{itemize}


\subsection{Prototype Configuration}\label{sec:configuration}

The prototype of \method's TD is configured with 32 vCPUs and 128 GB of memory, running Ubuntu 24.04.1 LTS with kernel version 6.8.0-86-generic~\cite{ubuntu_tdx}. The host has an Intel Xeon Silver 4514Y CPU, 512 GB of RAM, and an NVIDIA H100 NVL 94 GB GPU with confidential compute mode enabled~\cite{nvidia_guide}. We assign the GPU directly to the TD using Direct Device Assignment to give the TD exclusive access.
We run each experiment five times and report the average. Standard deviation is excluded as it was negligible.

\method is measurement/attestation is implemented in Python.
Attestation uses Intel Data Center Attestation Primitives (DCAP)~\cite{scarlata2018supporting} and NVIDIA Remote Attestation Service (NRAS)~\cite{nras}.
\method is built on PyTorch and the Hugging Face Datasets framework, allowing access to dataset and model internals for measurement. We use SHA-256 as the underlying hash function for normal hashing and \MSH. We use a 2048-bit prime for \MSH. For attestation, we use Intel DCAP to generate \texttt{QUOTE} and NRAS to verify $\gpuatt$. 
%
We also use the following operation-specific frameworks for the experiments:
\begin{itemize}[leftmargin=*]
    \item \textbf{Unsloth~\cite{unsloth}}: an open-source framework for optimized LLM fine-tuning. We use it for proof of fine-tuning experiment.
    \item \textbf{AutoAWQ framework~\cite{autoawq}}: implements the Activation-aware Weight Quantization (AWQ) algorithm for quantizing LLMs. We use it for proof of quantization experiment.
    \item \textbf{lm-evaluation-harness framework~\cite{eval-harness}}: standardized toolkit that runs LLMs on multiple benchmark tasks through a unified API to produce comparable scores. We use it for proof of evaluation, specifically evaluating the model with \overb{MMLU}.
\end{itemize}

\section{Evaluation}\label{sec:evaluation}


\subsection{Performance Analysis}\label{sec:efficiency}

Tables~\ref{tab:dataset_attestation}--\ref{tab:inference} show the timing results for the performance evaluations. The cells in \baseline{gray} represent the baseline, \method's overhead is highlighted in \overhead{blue}, and the total times are in \total{orange}. We parallelize the dataset lookup across 8 workers for dataset-related attestations requiring manual dataset iteration.


\subsubsection{Dataset Attestations}\label{sec:eval-dataset-att}
%
%
{
Table~\ref{tab:dataset_attestation} presents the results for proof of attribute distribution and proof of preprocessing. 
Both operations process the dataset using 8 parallel workers (i.e., vCPU cores). In the memory-mapped setting, parallel I/O can scale better with the memory-mapped approach, as all processes can directly access the same underlying memory pages without duplication.

As a result, the raw overhead (i.e., \MSH overhead per core) differs from the observed end-to-end overhead of the parallelized operation. The latter can be lower due to DataLoader prefetching~\cite{pytorch_data} or higher due to pipeline delays~\cite{arfeen2025pipefill, zhang2025freeride}. For example, raw overhead from \MSH operations of 62.35--68.06\% results in 37.64--42.50\% total observed overhead in the end-to-end parallelized operation. 

Despite the reduction in this case, the observed overhead remains high because \MSH requires modular exponentiation and multiplication over a 2048-bit prime for each record. Since dataset-level operations are inexpensive on per-record basis compared to others (e.g., fine-tuning, evaluation), even modest \MSH costs dominate the run-time. We analyze this effect further in Section~\ref{sec:mm-per-record-analysis}. 
%
Fortunately, these costs can be amortized across future dataset use. For example:
\begin{itemize}[leftmargin=*,nosep=*]
    \item proof of attribute distribution and preprocessing apply to all future uses of a dataset;
    \item proof of binding applies to any future dataset \MSH.
\end{itemize}
For the 1/100 subset of \overb{BookCorpus} dataset used in our evaluation, proof of binding takes 337.5 seconds (3.65 ms per record). 
}

\begin{table*}[!htb]
    \caption{Performance for proof of model optimization (fine-tuning and quantization).}
    \vspace{-1em}
    \centering
    \footnotesize
    \renewcommand{\arraystretch}{1.15}
    \begin{tabular}{l
        !{\vruletab}c!{\vruletab}c
        !{\dvruletab}c!{\vruletab}c
        !{\dvruletab}c!{\vruletab}c
        !{\dvruletab}c}

    \bottomrule

    \toprule
    \textbf{Proof of} &
    \multicolumn{6}{c!{\dvruletab}}{\textbf{Fine-tuning}} &
    \multicolumn{1}{c}{\textbf{Quantization}} \\
    \bottomrule

    \toprule


    \textbf{Model} &
    \multicolumn{2}{c!{\dvruletab}}{\bverb{Llama-3.1-8B}} &
    \multicolumn{2}{c!{\dvruletab}}{\bverb{Gemma-3-4B}} &
    \multicolumn{2}{c!{\dvruletab}}{\bverb{Phi-4-Mini}} &
    \multicolumn{1}{c}{\bverb{Llama-3.1-8B}} \\

    \textbf{Memory Variant} &
    \textbf{In-memory} & \textbf{Memory-mapped} &
    \textbf{In-memory} & \textbf{Memory-mapped} &
    \textbf{In-memory} & \textbf{Memory-mapped} &
    \textbf{N/A} \\
    \bottomrule

    \toprule
    \textbf{Total Time} &
    \totaltab{268.81 min} & \totaltab{261.74 min} &
    \totaltab{325.61 min} & \totaltab{369.56 min} &
    \totaltab{169.93 min} & \totaltab{170.48 min} &
    \totaltab{812.92 s} \\

    \midrule
    \textbf{Baseline} &
    \baselinetab{268.30 min} & \baselinetab{258.07 min} &
    \baselinetab{325.29 min} & \baselinetab{361.82 min} &
    \baselinetab{169.65 min} & \baselinetab{168.16 min} &
    \baselinetab{773.84 s} \\

    \textbf{Mem Usage (GB)} &
    \baselinetab{0.085} & \baselinetab{0.004} &
    \baselinetab{0.087} & \baselinetab{0.004} &
    \baselinetab{0.088} & \baselinetab{0.004} &
    \baselinetab{N/A} \\

    \midrule



    \textbf{Raw Meas.} &
    \overheadtab{28.76 s} & \overheadtab{132.87 s} &
    \overheadtab{17.93 s} & \overheadtab{129.40 s} &
    \overheadtab{15.83 s} & \overheadtab{127.99 s} &
    \overheadtab{28.16 s} \\

    \textbf{Attestation} &
    \overheadtab{1.11 s} & \overheadtab{1.44 s} &
    \overheadtab{1.00 s} & \overheadtab{1.20 s} &
    \overheadtab{1.00 s} & \overheadtab{0.84 s} &
    \overheadtab{0.92 s} \\

    \midrule
    \textbf{Raw Meas. Overhead (\%)} &
    \overheadtab{0.18} & \overheadtab{0.85} &
    \overheadtab{0.09} & \overheadtab{0.58} &
    \overheadtab{0.16} & \overheadtab{1.20} &
    \overheadtab{4.70} \\

    \textbf{Att. Overhead (\%)} &
    \overheadtab{0.007} & \overheadtab{0.009} &
    \overheadtab{0.005} & \overheadtab{0.005} &
    \overheadtab{0.010} & \overheadtab{0.008} &
    \overheadtab{0.117} \\
    
    \textbf{Total Observed Overhead (\%)} &
    \overheadtab{0.19} & \overheadtab{1.40} &
    \overheadtab{0.10} & \overheadtab{2.09} &
    \overheadtab{0.17} & \overheadtab{2.27} &
    \overheadtab{4.81} \\
    \bottomrule

    \toprule
    \end{tabular}
    \label{tab:finetune}
    \vspace{-1em}
\end{table*}

\begin{table*}[!htb]
    \caption{Performance for proof of evaluation  based on \overb{MMLU} and BLEU score with \overb{WMT14}\overb{(DE-EN)} dataset.}
    \vspace{-1em}
    \centering
    \LARGE
    \renewcommand{\arraystretch}{1.3}
    \setlength{\tabcolsep}{8pt}
    \resizebox{\textwidth}{!}{%
    \begin{tabular}{l
        !{\vruletab}c!{\vruletab}c!{\vruletab}c!{\vruletab}c
        !{\dvruletab}c!{\vruletab}c!{\vruletab}c!{\vruletab}c
        !{\dvruletab}c!{\vruletab}c!{\vruletab}c!{\vruletab}c}
    
    
    \bottomrule
        
    \toprule
   \textbf{Model}
        & \multicolumn{4}{c!{\dvruletab}}{\bverb{Llama-3.1-8B}}
        & \multicolumn{4}{c!{\dvruletab}}{\bverb{Gemma-3-4B}}
        & \multicolumn{4}{c}{\bverb{Phi-4-Mini}} \\

    \textbf{Memory Variant}
        & \multicolumn{2}{c!{\vruletab}}{\textbf{In-memory}}
        & \multicolumn{2}{c!{\dvruletab}}{\textbf{Memory-mapped}}
        & \multicolumn{2}{c!{\vruletab}}{\textbf{In-memory}}
        & \multicolumn{2}{c!{\dvruletab}}{\textbf{Memory-mapped}}
        & \multicolumn{2}{c!{\vruletab}}{\textbf{In-memory}}
        & \multicolumn{2}{c}{\textbf{Memory-mapped}} \\
     \bottomrule
        
    \toprule
    
    \textbf{Metric}
        & \overb{MMLU} & \textbf{BLEU} & \overb{MMLU} & \textbf{BLEU}
        & \overb{MMLU} & \textbf{BLEU} & \overb{MMLU} & \textbf{BLEU}
        & \overb{MMLU} & \textbf{BLEU} & \overb{MMLU} & \textbf{BLEU} \\
    \bottomrule
    
    \toprule
    \textbf{Total Time}
        & \totaltab{7.93 min} & \totaltab{121.71 min}
        & \totaltab{8.78 min} & \totaltab{121.55 min}
        & \totaltab{13.27 min} & \totaltab{242.23 min}
        & \totaltab{14.41 min} & \totaltab{241.53 min}
        & \totaltab{8.47 min} & \totaltab{119.03 min}
        & \totaltab{9.39 min} & \totaltab{118.65 min} \\
    \midrule
    \textbf{Baseline}
        & \baselinetab{7.73 min} & \baselinetab{121.23 min}
        & \baselinetab{7.80 min} & \baselinetab{120.94 min}
        & \baselinetab{13.15 min} & \baselinetab{241.92 min}
        & \baselinetab{13.30 min} & \baselinetab{241.10 min}
        & \baselinetab{8.35 min} & \baselinetab{118.75 min}
        & \baselinetab{8.43 min} & \baselinetab{118.26 min} \\
    \textbf{Mem Usage (GB)}
        & \baselinetab{0.028} & \baselinetab{0.001}
        & \baselinetab{0.013} & \baselinetab{0.001}
        & \baselinetab{0.022} & \baselinetab{0.001}
        & \baselinetab{0.008} & \baselinetab{0.001}
        & \baselinetab{0.025} & \baselinetab{0.001}
        & \baselinetab{0.009} & \baselinetab{0.001} \\
    \midrule
    \textbf{Raw Meas.}
        & \overheadtab{10.13 s} & \overheadtab{28.16 s}
        & \overheadtab{74.59 s} & \overheadtab{35.57 s}
        & \overheadtab{6.33 s} & \overheadtab{17.32 s}
        & \overheadtab{70.72 s} & \overheadtab{24.55 s}
        & \overheadtab{5.66 s} & \overheadtab{15.65 s}
        & \overheadtab{70.06 s} & \overheadtab{22.50 s} \\
    \textbf{Attestation}
        & \overheadtab{1.58 s} & \overheadtab{1.10 s}
        & \overheadtab{0.85 s} & \overheadtab{1.07 s}
        & \overheadtab{1.09 s} & \overheadtab{1.12 s}
        & \overheadtab{0.87 s} & \overheadtab{1.23 s}
        & \overheadtab{1.35 s} & \overheadtab{1.08 s}
        & \overheadtab{0.92 s} & \overheadtab{1.13 s} \\
    \midrule
    \textbf{Raw Meas. Overhead (\%)}
        & \overheadtab{2.13} & \overheadtab{0.39}
        & \overheadtab{14.16} & \overheadtab{0.49}
        & \overheadtab{0.79} & \overheadtab{0.12}
        & \overheadtab{8.18} & \overheadtab{0.17}
        & \overheadtab{1.11} & \overheadtab{0.22}
        & \overheadtab{12.43} & \overheadtab{0.31} \\
    
    \textbf{Att. Overhead (\%)}
        & \overheadtab{0.33} & \overheadtab{0.015}
        & \overheadtab{0.16} & \overheadtab{0.015}
        & \overheadtab{0.13} & \overheadtab{0.008}
        & \overheadtab{0.10} & \overheadtab{0.009}
        & \overheadtab{0.27} & \overheadtab{0.015}
        & \overheadtab{0.16} & \overheadtab{0.016} \\

    \textbf{Total Observed Overhead (\%)}
        & \overheadtab{2.46} & \overheadtab{0.40}
        & \overheadtab{11.24} & \overheadtab{0.50}
        & \overheadtab{0.93} & \overheadtab{0.13}
        & \overheadtab{7.69} & \overheadtab{0.17}
        & \overheadtab{1.38} & \overheadtab{0.23}
        & \overheadtab{10.28} & \overheadtab{0.31} \\
    \bottomrule
        
    \toprule
    \end{tabular}
    }
    \label{tab:eval}
\end{table*}

\subsubsection{Model Attestations}\label{sec:eval-model-att}
%
In contrast to dataset attestations, the measurement overhead of \method for proof of training is low. For the memory-mapped case, the raw measurement overhead is only 5.47\% of the total time, whereas for the in-memory case, it is 0.009\%; this results in an observed overhead of 5.61\% for memory-mapped and 0.014\% for in-memory case (Table~\ref{tab:pretrain}). LLM training is computationally expensive; hence, the measurement time introduces a \change{proportionally} smaller overhead. 



For model optimization, proof of fine-tuning (Table~\ref{tab:finetune}) also shows minimal overhead for both raw and observed of $\leq$2.27\% across all models, plus a substantial reduction in memory usage from 85-87 MB for the in-memory dataset to only 4 MB for the memory-mapped dataset.
For proof of quantization, since no dataset is involved, the 4.7\% raw measurement overhead comes from measuring models.


%
As shown in Table~\ref{tab:eval}, the \change{observed overhead of proof of evaluation with \overb{MMLU} is dominated by the measurement overhead}; the observed overheads are 0.93-2.46\% for the in-memory case and 7.69-11.24\% for the memory-mapped case (Table~\ref{tab:eval}). For BLEU score, since dataset access is limited (\overb{WMT14}\overb{(DE-EN)} contains only 3003 records), the raw overhead of performing \MSH in the memory-mapped case is comparable to hashing the dataset in the in-memory case, \edit{with both the raw and observed overheads remaining $\leq 0.5\%$} across all models.
\edit{Unlike dataset operations, total observed overhead for model operations is often higher than raw measurement overhead, likely due to pipeline stalls when transferring data to GPU.}



\subsubsection{Inference Attestations}\label{sec:eval-inf-att}
%
%
\begin{table*}[!htb]
    \caption{Performance of proof of inference for one query-response and proof of session inference for 50 query-responses.}
    \vspace{-1em}
    \centering
    \footnotesize
    \renewcommand{\arraystretch}{1.1}
    \setlength{\tabcolsep}{14pt}
    \begin{tabular}{l!{\vruletab}c!{\vruletab}c!{\dvruletab}c!{\vruletab}c!{\dvruletab}c!{\vruletab}c}
    \bottomrule
    
    \toprule
    \textbf{Model} 
        & \multicolumn{2}{c!{\dvruletab}}{\bverb{Llama-3.1-8B}} 
        & \multicolumn{2}{c!{\dvruletab}}{\bverb{Gemma-3-4B}} 
        & \multicolumn{2}{c}{\bverb{Phi-4-Mini}} \\
    \textbf{Proof of Inference Type} 
        & \textbf{Single} & \textbf{Session} 
        & \textbf{Single} & \textbf{Session} 
        & \textbf{Single} & \textbf{Session} \\
    \bottomrule
    
    \toprule
    \textbf{Total Time} 
        & \totaltab{43.31 s} & \totaltab{256.19 s} 
        & \totaltab{40.55 s} & \totaltab{489.69 s} 
        & \totaltab{29.94 s} & \totaltab{249.37 s} \\
    \midrule
    \textbf{Baseline} 
        & \baselinetab{14.54 s} & \baselinetab{227.02 s} 
        & \baselinetab{22.09 s} & \baselinetab{471.22 s} 
        & \baselinetab{13.40 s} & \baselinetab{232.70 s} \\
    \textbf{Mem Usage (GB)}
        & \baselinetab{0.001} & \baselinetab{0.001}
        & \baselinetab{0.001} & \baselinetab{0.001}
        & \baselinetab{0.001} & \baselinetab{0.001} \\
    \midrule
    \textbf{Raw Meas.} 
        & \overheadtab{27.17 s} & \overheadtab{28.25 s} 
        & \overheadtab{16.27 s} & \overheadtab{17.47 s} 
        & \overheadtab{14.59 s} & \overheadtab{15.66 s} \\
    \textbf{Attestation} 
        & \overheadtab{0.75 s} & \overheadtab{0.92 s} 
        & \overheadtab{0.79 s} & \overheadtab{0.99 s} 
        & \overheadtab{0.71 s} & \overheadtab{1.01 s} \\
    \midrule
    \textbf{Raw Meas. Overhead (\%)} 
        & \overheadtab{64.34} & \overheadtab{11.03} 
        & \overheadtab{43.28} & \overheadtab{3.57} 
        & \overheadtab{52.16} & \overheadtab{6.28} \\
    \textbf{Att. Overhead (\%)} 
        & \overheadtab{2.09} & \overheadtab{0.36} 
        & \overheadtab{2.24} & \overheadtab{0.20} 
        & \overheadtab{3.10} & \overheadtab{0.41} \\
    \textbf{Total Observed Overhead (\%)} 
        & \overheadtab{66.43} & \overheadtab{11.39} 
        & \overheadtab{45.52} & \overheadtab{3.77} 
        & \overheadtab{55.26} & \overheadtab{6.69} \\
    \bottomrule
    
    \toprule
    \end{tabular}
    \label{tab:inference}
\end{table*}

Table~\ref{tab:inference} presents the results for both proof of inference and proof of session inference.
Proof of inference follows the same pattern observed in dataset attestations, where the overhead appears large (64.34\% for \bverb{Llama-3.1-8B}, 43.28\% for \bverb{Gemma-3-4B}, and 52.16\% for \bverb{Phi-4-Mini}) because the underlying computation is relatively small. This represents an uncommon case in which a proof is generated for only one prompt.
On the other hand, 
proof of session inference simulates a more realistic user-model interaction.
In this setting, the measurement overhead is significantly smaller: 11.03\% for \bverb{Llama-3.1-8B}, 3.57\% for \bverb{Gemma-3-4B}, and 6.28\% for \bverb{Phi-4-Mini}.

\subsubsection{Measurement overhead of memory-mapped operations}\label{sec:mm-per-record-analysis}

Table~\ref{tab:per-record} compares per-record computation time with the \MSH construction time (i.e., input measurement) for operations that use a memory-mapped dataset.
Normalizing per record shows that dataset-related operations are much more lightweight than model-related operations.
In contrast, input measurement time per record remains nearly constant ($\approx$2.29--2.57 ms), as it is independent of the operation type. 
Consequently, relative measurement overhead is higher for operations with lower computational cost, despite the near-constant per-record measurement time.
Due to the low overhead for common operations (training, fine-tuning, quantization, evaluation, session inference), 
\method satisfies \textbf{R1}.

\begin{table}[t]
    \caption{Records per workers (R/W), compute time per record (C/R), and raw input measurement time per record (I/R) for memory-mapped operations. I/R remains roughly constant across operations, while C/R varies significantly, indicating that raw measurement overhead is higher when C/R is low.}\label{tab:per-record}
    \vspace{-1em}
    \centering
    \renewcommand{\arraystretch}{1.2}
    \resizebox{\columnwidth}{!}{
    \LARGE
    \begin{tabular}{l!{\vruletab}c!{\vruletab}c!{\vruletab}c!{\vruletab}c!{\dvruletab}c}
    \bottomrule
        
    \toprule
    \shortstack{\textbf{Operation}\\~} & \shortstack{\textbf{Model}\\~} & \shortstack{$\mathbf{R/W}$\\~} & \shortstack{$\mathbf{C/R}$ \textbf{(ms)}\\~} & \shortstack{$\mathbf{I/R}$ \textbf{(ms)}\\~} & \shortstack{\textbf{Raw Meas.} \\ \textbf{Overhead (\%)}} \\
     \bottomrule
        
    \toprule
         \textbf{Training} & \bverb{GPT2} & 572.2k & 43.07 & 2.44 & 5.47 \\
    \midrule
         \textbf{Fine-tuning} & \bverb{Llama-3.1-8B} & 51.8k & 311.1 & 2.36 & 0.84 \\
         \textbf{Fine-tuning} & \bverb{Gemma-3-4B} & 51.8k & 376.3 & 2.38 & 0.58 \\
         \textbf{Fine-tuning} & \bverb{Phi-4-Mini} & 51.8k & 196.9 & 2.36 & 1.20 \\
    \midrule
         \textbf{Eval.} (\overb{MMLU}) & \bverb{Llama-3.1-8B} & 28.1k & 40.5 & 2.29 & 14.15 \\
         \textbf{Eval.} (\overb{MMLU}) & \bverb{Gemma-3-4B} & 28.1k & 33.8 & 2.29 & 8.17 \\
         \textbf{Eval.} (\overb{MMLU}) & \bverb{Phi-4-Mini} & 28.1k & 27.4 & 2.29 & 12.43 \\
    \midrule
         \textbf{Eval. (BLEU)} & \bverb{Llama-3.1-8B} & 3.0k & 2416 & 2.34 & 0.45 \\
         \textbf{Eval. (BLEU)} & \bverb{Gemma-3-4B} & 3.0k & 4817.5 & 2.35 & 0.14 \\
         \textbf{Eval. (BLEU)} & \bverb{Phi-4-Mini} & 3.0k & 2363.7 & 2.34 & 0.30 \\
    \midrule
         \textbf{Attr. Dist.} & N/A & 9.25M & 3.75 & 2.54 & 68.06 \\
         \textbf{Preprocessing} & N/A & 92.5k & 4.09 & 2.57 & 62.35 \\
    \bottomrule
    
    \toprule
    \end{tabular}
    }
\end{table}

\subsection{Security Analysis}\label{sec:security}


\method aims to uphold integrity of its property attestation evidence in the presence of \adv, as defined in Section~\ref{sec:problem}, who controls the host machine, including the VMM and disk. Based on this, the following security properties are desired:
\begin{enumerate}[label=\textbf{[P\arabic*]},nosep]
    \item integrity of the operation (the set of $\mathcal{H_{I}}$ and $\mathcal{H_O}$ correspond to the expected $\mathcal{I}, \mathcal{O}$, according to the requested $op$).
    \item integrity of environment and responses ($h_{TD}$, $op$, $\mathcal{H_{I}}$, and \change{$\mathcal{H_{O}}$} in \quote correspond to the expected TD image, \change{$\mathcal{I}$, op, and $\mathcal{O}$}, respectively);
    \item authenticity of attestation responses (any \quote that passes verification was produced through the expected QE, and \tdreport was not forged); and
\end{enumerate}
\change{In order to violate these security properties}, \adv attempts to
(1) tamper with any operation inputs; 
(2) tamper with, discard, or replay \tdreport or \quote at any stage of their construction or transmission;
or (3) tamper with the configuration of \method TD.
To verify that the property attestation protocol from Section~\ref{sec:attestation} upholds \textbf{[P1-3]}, we use the Tamarin Prover symbolic verification tool~\cite{tamarin} to model the protocol depicted in Figure~\ref{fig:protocol}, to construct a set of security goals that subsume \textbf{[P1-3]}, and to verify that the security goals are upheld. Here, we describe the rationale for why they are upheld, with complete details in Appendix~\ref{apdx:tamarin}.

\adv may tamper with operation inputs to \change{violate any of these properties}. For example, \adv may manipulate $op$ to change the requested function or misconfigure assets in $\mathcal{I}$ used for the requested function. Both cases are detectable by \vrf, since \method does not measure either until they are loaded into TD memory. Therefore, \vrf is \change{assured} that the inputs reflected in the \quote were used in the process. In the case of a memory-mapped data structure, \adv may manipulate data before it is read into the TD memory. For example, they may aim to repeat, replace, or reorder data records during fine-tuning to \change{poison or bias} the resulting model. These attempts can be detected by \vrf, since \method measures a \MSH of records and indices as they are sampled. We use Lemmas
~\ref{tamarin:lemma:key_sec}, 
~\ref{tamarin:lemma:o_from_op}, 
~\ref{tamarin:lemma:input_integrity}, 
~\ref{tamarin:lemma:input_integrity_mm}, 
~\ref{tamarin:lemma:msh-implies-multiset}, 
~\ref{tamarin:lemma:ms_hash_match}, 
~\ref{tamarin:lemma:msh_o_match}, 
and~\ref{tamarin:lemma:ms_O_match} 
to verify the properties are upheld despite these attempts.

\adv may attempt to generate false outputs and corresponding measurements in an attempt to create a faulty \tdreport before it is passed to the QE. 
\edit{\method relies on the standard TDX attestation guarantees discussed in Section~\ref{sec:background} that (a) \tdreport authenticity is protected by MAC under a hardware-protected key inaccessible to software~\cite{tdx_security_report} and (b) the QE verifies this MAC prior to the \quote generation.}
\edit{ Consequently, \adv cannot forge a valid \tdreport without access to the device key, and \quote forgery is infeasible because each \quote is signed using a provisioned attestation key whose private material never leaves QE.}
\edit{\method only extends data into \texttt{REPORTDATA} of the \tdreport. Since material extended into \texttt{REPORTDATA} by \method is independent of keys, and \method does not influence any other step the \quote generation pipeline, the integrity/authentication guarantees of Intel TDX attestation remain unchanged.}
\adv may attempt to replay responses from the same $op$ to \vrf, but this is detectable by \vrf due to \method incorporating $\chal$ as input in construction of \tdreport (and subsequently \quote). 
We use Lemmas~\ref{tamarin:lemma:key_sec}, 
~\ref{tamarin:lemma:quote_authenticity}, 
~\ref{tamarin:lemma:quote_gen_if_mac}, 
~\ref{tamarin:lemma:input_integrity}, 
and~\ref{tamarin:lemma:input_integrity_mm} 
to ensure that a verifiable \quote was produced after an integrity check on a MAC-ed \tdreport, and no secret keys are revealed through the protocol.

Since \adv has control over \change{the} VMM, they can assign which image is loaded into the TD. Therefore, they could load an image containing a backdoor to assist with lying about datasets/models that are used for a particular operation. However, the TD image is measured by the TDX module before TD creation, and included in the \quote obtained by \vrf. Therefore, these attempts are detectable by \vrf.
Furthermore, \adv may attempt to assign a GPU without a TEE to the TD. In this case, \vrf detects it due to \method including GPU attestation evidence when the GPU is used in obtaining outputs. In this case, \vrf would observe $\gpuatt$'s absence from the response.
We use Lemmas~\ref{tamarin:lemma:key_sec} and~\ref{tamarin:lemma:td_integrity} to show that a \quote containing \method configuration (e.g., TD image and GPU) is not accepted by \vrf unless the reporting TD was created with the same configuration.
We do not explicitly model the scenario of GPU reassignment in our formal verification: we assume the SPDM key exchange between the GPU and TD occurs before TD launch, as it is impossible to launch a GPU in CC-mode without this~\cite{gu2025nvidia,spdm}. Given this, we assume the GPU–TD pair is a single trust boundary in our model.

Finally, when outputs are deemed private, \adv may try to corrupt \method TD to reveal them. However, through \method measurer script and construction of \tdreport, confidential outputs themselves are never extended into any \tdreport or any downstream \quote that are visible by \adv. We use Lemmas~\ref{tamarin:lemma:private_o_inmem} and~\ref{tamarin:lemma:private_o_mem_mapped} outlined in outlined in Appendix~\ref{apdx:tamarin} to verify that this is upheld.


Under the assumed threat model, \adv cannot tamper with operation inputs, forge or replay attestation responses, or manipulate the TD configuration without detection by \vrf. Furthermore, the property attestation protocol is formally verified to uphold \textbf{[P1-3]} despite these attempts, as confirmed by the Tamarin model detailed in Appendix~\ref{apdx:tamarin}. Therefore, \method satisfies \textbf{R2}.

\subsection{Scalability \& Versatility}\label{sec:others}

\textbf{Scalability.} 
\method satisfies \textbf{R3} as it can be applied to extremely large datasets and models, distributed computing by \prv, and enables verification by multiple verifiers.

\change{\method's use of \MSH allows it to scale to extremely large datasets, beyond what can fit into memory, without sacrificing out-of-order access}. We assume model weights can fit into the GPU's RAM (e.g., 96 GB for our prototype); when they do not, standard optimizations (e.g., quantization) reduce memory impact while maintaining performance. While some ML frameworks support memory-mapped model weights~\cite{llamacpp_model_mem_map,hf_model_mem_map}, accessing them through GPU$\leftrightarrow$CVM$\leftrightarrow$disk adds significant overhead. Recent work shows quantization can eliminate the need for weight offloading~\cite{flexgen}. \change{When quantization is insufficient or undesirable}, \MSH can also apply to memory-mapped model weights, \change{and not just datasets}.

\method can extend to distributed training~\cite{distributed_training}, where each node returns proof(s) for its dataset subset. Due to \MSH’s incremental design, it enables aggregation of component hashes into that of the full dataset. However, verification complexity, node-level corruption, and Byzantine resilience require further study, leaving \textit{proofs of distributed training} as an intriguing direction for future work.

\method uses digital signatures to create each \texttt{QUOTE} containing the property attestation. Therefore, any number of verifiers with access to  the attesting device's public key can perform verification.


\textbf{Versatility.}
\method satisfies \textbf{R4} as it can apply to any CVM and GPU with TEE, and it can be adapted to any type of model architecture or dataset.
Although \method prototype uses Intel TDX as a CVM and NVIDIA H100 for TEE-enabled GPU, \method's attestations can be generated by \prv running on 
any CVM-GPU configuration, given both have TEEs with built-in attestation functionality (e.g., AMD SEV-SNP~\cite{amd-sev-snp} and NVIDIA Blackwell GPU~\cite{nvidia_cc}).
%
Furthermore, \method can be adapted to any type of model architecture or dataset, beyond those used in our experiments, as measurements are specified in software running inside a CVM that invokes a GPU with a TEE. Property attestations can also be easily added by extending \method with additional measurer scripts corresponding to the new property, with no changes required by \intr or \vrf. 


\section{Related Work}\label{sec:related}


\textbf{Cryptographic Attestations.}
Secure multi-party computation (SMPC) and zero-knowledge proofs (ZKPs) have been used to verify ML operations.
SMPC has been explored for distributional property attestation but requires per-attestation interaction with a verifier, which is not scalable~\cite{duddu2024attesting}. Arc uses SMPC with publicly verifiable commitments to bind training and inference for auditing~\cite{lycklama2024holding}.
ZKPs can provide verifiability for training simple ML models (e.g., logistic regression)~\cite{zkpot1} and neural networks~\cite{zkpot2}, and LLM fine-tuning~\cite{zklora}.
ZKPs have also been used for proof of inference across different types of ML models (including LLMs)~\cite{zkmlaas,zkcnn,zkllm}.
Furthermore, ZKPs can help verify security properties of ML models such as differential privacy~\cite{shamsabadi2024confidentialdpproof,cryptoeprintFranzese}, and fairness~\cite{franzese2024oath,shamsabadi2023confidentialprofitt}.
However, these cryptographic primitives have a high computation cost. For instance, verifying one iteration of VGG11 training takes $\approx$15 minutes~\cite{zkpot2}, making practical deployment challenging for applications.

\textbf{TEE-based attestations.}
Torres-Arias et al.~\cite{236322} were the first to introduce supply chain attestations. However, their work targets conventional software pipelines rather than ML-specific properties.
Prior works have extensively used TEEs for the confidentiality of data and models~\cite{teeslice,tensorshield,mirrornet,darknetz}.
Unlike these approaches, we focus on integrity, leveraging TEEs to attest properties of large generative model operations executed within them.
Laminator~\cite{duddu2024laminator} is the closest related work, using Intel SGX to attest properties of CPU-only classifier model operations.
Schnabl et al.~\cite{attestable_audits} use TEEs to attest the execution of the code, ML model for various inferences, and its outputs. 
\emph{VerifiableFL} proposes an integrity-only enclave (called \emph{exclave}), which generates authenticated evidence of input and output data for federated learning~\cite{exclavefl}. Similar to our approach, it uses CVM technologies (e.g., AMD SEV-SNP), but focuses on federated learning rather than generative models, and does not focus on addressing the challenges due large datasets outside CVM memory. 
Rattanavipanon and Nunes~\cite{slapp} focus on federated learning by providing stateful proofs of execution that let an external \vrf confirm correct data use, model updates, and local differential privacy on TrustZone-enabled ARM Cortex-M devices.
\emph{Atlas} uses TEEs to verify the ML lifecycle by having TEE-equipped provers attest to inputs and outputs of artifact operations and publish attestations to a transparency log~\cite{atlas}.

None of the mentioned works focus on attesting properties of generative models while accounting for challenges of their computing environments outlined in Section~\ref{sec:problem}.
Most closely related is Atlas, yet it differs from \method in its attestation semantics. Atlas proves that artifacts follow expected pipeline steps, whereas \method proves that outputs satisfy operation-level properties for given inputs. Atlas excludes GPU execution from its TCB and uses whole-file hashing, which introduces TOCTOU risks for dynamically sampled datasets. \method addresses these with the use of TEE-aware GPU and \MSH when sampling from untrusted storage.

\section{Summary}
This work advances ML property attestations by introducing \method: a comprehensive framework capable of spanning CPU-GPU environments, large-scale data, and emerging generative models. We define how to measure properties of generative models and use \textit{incremental multiset hash} functions to capture properties across dataset operations and training. As a result, \method enables accountability for modern AI systems, laying the groundwork for regulators, developers, and verifiers to concretely reason about an ML systems' trustworthiness.
\clearpage


\bibliographystyle{ACM-Reference-Format}
\bibliography{paperL}
\appendix
\section*{Appendix}


\section{Additional Considerations}\label{sec:discussions}

\textbf{Other Generative Models.}
Since \method is built on PyTorch and Hugging Face, it applies to any hosted model with accessible weights, including other generative models not evaluated in this work (e.g., diffusion models~\cite{HoJA20}).
For diffusion models, only the measured input and output changes (e.g., images instead of text). Properties may differ and can be added due to \method versatility, however the approach for attestation would remain unchanged.


\textbf{Other Data Integrity Techniques.}
\method uses \MSH to incrementally hash dataset records into a single value, independent of the access order, but requires all record are read exactly once.
Thus, it applies only to full-dataset operations (common in LLM workloads). 
If partial access within a dataset is required, a Merkle tree can be used at an $\mathcal{O}(\log n)$ cost per lookup. 
A \textit{proof of binding} can link \MSH to a Merkle root, enabling selection based on system requirements. 

\textbf{Selecting \chal.} Section~\ref{sec:attestation} outlines a property attestation protocol enabled by \method initiated by request from \intr containing \chal. In conventional remote attestation settings, \chal is a nonce or counter used to prevent replay attacks and ensure response freshness. Given ML property attestation in this work is non-interactive with \vrf, selection of \chal should be derived from a timestamp rather than a traditional cryptographic nonce or counter. 
However, for proof of inference, a nonce remains valuable to prevent \prv from cherry-picking favorable executions. In this case, \vrf can publish nonce values with expiration times for use challenges. To further mitigate cherry-picking, \vrf may request only proof of session inference, allowing observability of response history.

\section{Overhead of Elliptic Curve-based and Discrete Logarithm-based Construction}\label{apdx:msh-comparison}

The \MSH overhead observed in Section~\ref{sec:efficiency} stems from the hardness of the discrete logarithm problem, which requires operating over a large prime (in our case, a 2048-bit prime). Prior work has explored elliptic curve-based constructions~\cite{eech, ecmh}. Maitin-Shepard, Tibouchi, and Aranha proposed \texttt{ECMH}~\cite{ecmh}, a \texttt{Mset-Mu-Hash} scheme based on binary elliptic curves and characteristic-2 variant of the Shallue-van de Woestijne encoding~\cite{sw06} for deterministic hashing to curve points. A simpler alternative is the try-and-increment method~\cite{boneh2001short}, which repeatedly attempts to map a value onto a valid elliptic curve point until successful.

We test proof of attribute distribution using an elliptic-curve-based \MSH with try-and-increment over the NIST P-256 curve, and we observe a reduction in raw measurement overhead by 47.41\%. This further supports the analysis from Section~\ref{sec:eval-dataset-att} that the dominant source of overhead is the \MSH construction itself.
%
Because try-and-increment has input-dependent timing~\cite{rfc9380}, it is suitable only for public datasets. For private datasets, deterministic constructions, such as \texttt{ECMH}, should be used. 







\section{Operations and Properties in \method}\label{apdx:property-table}

Table~\ref{tab:op-to-prop} shows operations and resulting property attestations.

\begin{table}[t]
    \centering
    \caption{Operations and property attestations in \method}
    \label{tab:op-to-prop}
    \vspace{-1em}
    \footnotesize
        \resizebox{0.85\columnwidth}{!}{
    \begin{tabular}{{l!{\vruletab}l}}
    \bottomrule
    
    \toprule
        \textbf{Operation ($op$)} & \textbf{Property Attestation} \\
    \midrule
        Preprocessing & Proof of preprocessing\\

        Attribute distribution & Proof of attribute distribution\\

        Measurement binding & Proof of binding \\

        Training & Proof of training \\

        Weight optimization & Proof of optimization\\

        Fine-tuning & Proof of fine-tuning\\

        Quantization & Proof of quantization \\

        Evaluation & Proof of evaluation \\

        Inference & Proof of inference \\

        Chat session inference & Proof of session inference \\
    \bottomrule
    
    \toprule
    \end{tabular}
    }
\end{table}

\section{Formal Modeling with Tamarin}\label{apdx:tamarin}
To prove the security properties of~\method, we model its property attestation protocol (from Figure~\ref{fig:protocol}) in the Tamarin Prover~\cite{tamarin}. Tamarin is a symbolic protocol verification tool. Protocols and adversaries are modeled as multiset rewrite rules with preconditions and postconditions (called \textit{facts}). Security goals are specified as \textit{lemmas}, which Tamarin verifies by matching rule preconditions and postconditions.

\begin{figure}[ht]
\begin{lstlisting}[style=tamarinstyle]
|\ruleheader{Example Rule}\label{tamarin:rule:example}|
rule Example_Rule:
  [ Fr(~A), In(B)] 
  --[Action(~A,B)]->
  [State(~A,B), Out(h(<~A,B>)]
\end{lstlisting}
\end{figure}

Consider the example in Rule~\ref{tamarin:rule:example}; \textit{Example\_Rule} has two preconditions: \textit{Fr(\textasciitilde{}A)} and \textit{In(B)}. It can only be evaluated if a \textit{fresh} symbol $A$ exists (signaled to Tamarin using \textit{Fr(\textasciitilde{})}), and if a symbol $B$ was received from an \adv-controlled channel (e.g., the network).

This rule creates two postconditions: \textit{State(\textasciitilde{}A, B)}, that can be used as a precondition in other rules, and  \textit{Out(h(<\textasciitilde{}A, B>))} which sends \textit{h(<\textasciitilde{}A, B>)} to an \adv-controlled channel.
This example uses the \textit{hashing} built-in, which provides a representation of a cryptographic hash function \textit{h()}, to output the hash of \textit{A} and \textit{B} concatenated.
We use \textit{h()} in our model, along with \textit{signing} built-in, which provides digital signature functions \textit{sign()} and \textit{verify()}. 

An \textit{action fact} records that a rule was evaluated. In Rule~\ref{tamarin:rule:example}, the action fact \textit{Action(\textasciitilde{}A, B)} records that this rule was evaluated with symbols \textit{A} and \textit{B}. Lemmas in Tamarin must be specified using \textit{symbols} and \textit{action facts}, not precondition or postcondition facts.

\begin{figure}[ht]
\begin{lstlisting}[style=tamarinstyle]
|\lemmaheader{Example Lemma}\label{tamarin:lemma:example}|
lemma Example_Lemma:
  "All A B #i.
  Action(A,B) @i ==> not (Ex #j. K(A) @ #j)
\end{lstlisting}
\end{figure}

Lemma~\ref{tamarin:lemma:example} is an example lemma \textit{Example\_Lemma} using the symbols: 
\textit{A}, \textit{B}, and \textit{\#i}.
The notation \textit{\#i} tells Tamarin that the symbol \textit{i} is a time variable. 
This lemma states: given \textit{Action(A,B)} occurs at time \textit{\#i}, then there does not exist a time \textit{\#j} in which \adv learns \textit{A}. 
\adv's knowledge is expressed using the built-in fact $K()$.

Assuming Rule~\ref{tamarin:rule:example} is the only rule, Tamarin evaluates explores all possible executions, controlling inputs via \textit{In()} and outputs via \textit{Out()} to attempt to learn \textit{A}. This lemma is \textit{verified} no execution exists in which \textit{A} is learned.

In our setting, QE, the TDX module, and \method TD are trusted.
Communication between the TDX module, \method TD, and GPU occurs over secure channels.
This manifests as custom precondition/postcondition facts.
Channels that traverse the VMM, disk, or the network are considered untrusted. For example, QE-TDX Module communication passes through the VMM, dataset inputs are read from disk during mem-mapped operation, and each \quote is transmitted over the network prior to verification.
These interactions are modeled using \textit{In} and \textit{Out} facts.

\subsection{Security Goals}\label{appdx:tamarin_sec_goal}


Recall from Section~\ref{sec:security} the properties \textbf{[P1-3]}. We formalize them as the following goals that we aim to prove in our formal model:
\begin{enumerate}[leftmargin=*,label=\textbf{[G\arabic*]},nosep]
    \item \textbf{Confidentiality of Secrets}: \method measurements and outputs should not reveal secrets used to generate a \quote;
    
    \item \textbf{TD Measurement Integrity:} a \quote is only valid if produced from a valid TD image that was indeed installed;
    
    \item \textbf{Quote Authenticity:} any accepted \quote is signed by QE using its private key;
    
    \item \textbf{Quote Integrity:} any accepted \quote was produced after performing a MAC integrity check on a \texttt{TDREPORT};
    
    \item \textbf{Output Integrity}: any accepted \quote uses $\mathcal{O}$ that was produced via $op$ execution over specified inputs $\mathcal{I}$; and
    
    \item \textbf{Input Measurement Integrity:} any accepted \quote was produced using $\mathcal{I}$ that was loaded into TD-protected memory. 
\end{enumerate}

\begin{figure}[h]
\centering
\smaller
\fbox{
    \parbox{0.95\columnwidth}{
        \textbf{\underline{Properties from Goals:}}
        \begin{equation*}
        \begin{aligned}
            \textbf{[P1]} &\leftarrow \textbf{[G1]} \land \textbf{[G5]} \land \textbf{[G6]} \\
            \textbf{[P2]} &\leftarrow \textbf{[G1]} \land \textbf{[G2]} \land \textbf{[G5]} \land \textbf{[G6]} \\
            \textbf{[P3]} &\leftarrow \textbf{[G1]} \land \textbf{[G3]} \land \textbf{[G4]}
        \end{aligned}
        \end{equation*}
}
}
\end{figure}

\textbf{[P1]} concerns operation integrity. It is established by \textbf{[G1]}, \textbf{[G5]}, and \textbf{[G6]}. 
\textbf{[G1]} ensures MAC/signature keys are not leaked by the protocol.
\textbf{[G5]} and \textbf{[G6]} ensure that output/input measurements correspond to the operation.

\textbf{[P2]} concerns integrity of the environment and responses. It is established by \textbf{[G1]}, \textbf{[G2]}, \textbf{[G5]}, and \textbf{[G6]}. 
\textbf{[G2]}, \textbf{[G5]}, and \textbf{[G6]} ensure response integrity, as it includes $h_{TD}$, $\mathcal{H_O}$, and $\mathcal{H_I}$.

\textbf{[P3]} concerns response authenticity. It is established by \textbf{[G1]}, \textbf{[G3]}, and \textbf{[G4]}. \textbf{[G1]} again ensures key secrecy, while \textbf{[G3]} and \textbf{[G4]} ensure any verifiable \quote was signed by the expected QE and MAC-ed by the expected TDX machine, respectively. 

We use a Tamarin model to evaluate \textbf{[G1-6]} under four protocol scenarios, varying by dataset memory model and privacy of operation outputs:
\begin{itemize}[leftmargin=*,,nosep]
    \item in-memory datasets to produce public outputs (\textbf{Case 1});
    \item in-memory datasets to produce private outputs (\textbf{Case 2});
    \item memory-mapped datasets to produce public outputs (\textbf{Case 3});
    \item memory-mapped datasets to produce private outputs (\textbf{Case 4}).
\end{itemize}

\begin{figure}[t]
\begin{lstlisting}[style=tamarinstyle]
// Case 1: In-memory and public outputs
|\ruleheader{TD Initialization}\label{tamarin:rule:init_td}|
rule Init_TDX_TD:
  [ Fr(~k), Fr(~qsk), In(tdImage) ] 
  --[ InitTD(tdImage), SecretKey(~k), SecretKey(~qsk), InitialInstall(tdImage) ]->
  [ !Key(~k), !QESignKey(~qsk), !QEVerKey(pk(~qsk)), Out(pk(~qsk)), !TDIsRunning(tdImage) ]

|\ruleheader{Get Inputs}\label{tamarin:rule:get_inputs_inmem_public}|
rule Get_Operation_Inputs:
  [ In(op), In(I), In(chal), !TDIsRunning(hTD)] 
  --[ InputMessage(op, I, chal) ]->
  [ OperationInputs(op, I, chal, hTD)]

|\ruleheader{Operation Execution (In-memory)}\label{tamarin:rule:op_exec_inmem_public}|
rule Execute_Operation:
  [ OperationInputs(op, I, chal, hTD) ]
  --[ RunOperation(op, I, chal, runOp(op, I)) ]->
  [ CompleteOperation(op, I, chal, runOp(op, I), hTD) ]

|\ruleheader{TD Report Generation}\label{tamarin:rule:tdreport_gen_inmem_public}|
rule TD_Produce_Report:
    let
      O = runOp(op, I)
      mac = h(<h(I), h(op), h(chal), h(O), hTD, k>)
    in
    [ !Key(k), CompleteOperation(op, I, chal, runOp(op, I), hTD) ]
  --[ TDProducesReport(op, I, chal, runOp(op, I), hTD, k) ]->
    [ Out(<O, h(I), h(op), h(chal), h(O), hTD, mac>), ReportReady(op, I, chal, runOp(op, I), hTD, k) ]

|\ruleheader{Quote Generation}\label{tamarin:rule:quote_gen_inmem_public}|
rule QE_Produce_Quote:
    let 
      mac = h(<hI, hOp, hChal, hO, hTD, k>)
      sig = sign(<hI, hOp, hChal, hO, hTD>, qsk)
      O = runOp(op, I)
    in
    [ !QESignKey(qsk), In(<O, hI, hOp, hChal, hO, hTD, mac>), ReportReady(op, I, chal, O, hTD, k) ]
  --[ MacCheck(hI, hOp, hChal, hO, hTD, mac, k),
      QEQuotes(hI, hOp, hChal, hO, hTD, pk(qsk)) ]->
    [ Out(<O, sig>), SignatureReady(op, I, chal, O, hTD, sig) ]

|\ruleheader{Verification}\label{tamarin:rule:quote_ver_inmem_public}|
rule Vrf_Verifies:
    [ !QEVerKey(pkQ),
      SignatureReady(op, I, chal, O, hTD, sig),
      In(<O, sig>) 
    ]
  --[ VrfAccepts(op, I, chal, O, hTD, verify(sig, <h(I), h(op), h(chal), h(O), hTD>, pkQ), pkQ) ]->
    [] 
\end{lstlisting}
\end{figure}

\subsection{Case 1: In-Memory Public Outputs}\label{appdx:tamarin-case1}

We divide the protocol for Case 1 into three parts: (i) TD initialization, (ii) operation execution, and (iii) \tdreport generation, \quote signing, and verification.

\textbf{TD Initialization.} We model steps (1)–-(2) of the TD initialization phase from Figure~\ref{fig:protocol} using Rule~\ref{tamarin:rule:init_td}. This rule models TD creation (with GPU assigned and $GPU_{att}$ obtained), provisioning cryptographic keys, and the initial state measurement.

This rule takes two fresh secret values, an internal MAC key (${\sim}k$) and a QE signing key (${\sim}qsk$), and a TD image as inputs, producing persistent facts for the resulting TD state: the MAC key, the QE signing/verification key pair, and a running TD loaded with that image. The QE public key is output via the untrusted channel to \adv, while private keys are logged via \textit{SecretKey} action facts. The \textit{InitTD} and \textit{InitialInstall} action facts capture the CPU measurement and extension of the TD's initial software state.

\textbf{Operation Execution.}
We model steps (1)--(5) after TD initialization from Figure~\ref{fig:protocol} with two rules: (i) fetch operation inputs (e.g., $op$, $\chal$, and $\mathcal{I}$) in Rule~\ref{tamarin:rule:get_inputs_inmem_public} and (ii) operation execution (Rule~\ref{tamarin:rule:op_exec_inmem_public}).

Rule~\ref{tamarin:rule:get_inputs_inmem_public} consumes an $op$, $\mathcal{I}$, and $\chal$ from the network alongside a running TD fact, binding them together into an~\textit{OperationInputs} fact that scopes execution to a live TD. Rule~\ref{tamarin:rule:op_exec_inmem_public} consumes these inputs and applies the abstract function~\textit{runOp(op, I)} which computes the operation output. Then, we log the execution via the~\textit{RunOperation} action fact and forward all relevant values as a~\textit{CompleteOperation} fact for the next steps.

\textbf{\tdreport/\quote Generation and Verification}.
This part models steps (6)--(11) of Figure~\ref{fig:protocol}. Rule~\ref{tamarin:rule:tdreport_gen_inmem_public} models \tdreport generation. 
Then, Rule~\ref{tamarin:rule:quote_gen_inmem_public} models transferring \tdreport to QE for signing via (untrusted) VMM. Lastly, the Rule~\ref{tamarin:rule:quote_ver_inmem_public} models making signed \quote available to \vrf for verification.

In Rule~\ref{tamarin:rule:tdreport_gen_inmem_public}, TD computes a MAC over the hashes of $\mathcal{I}$, $op$, $\chal$, $\mathcal{O}$, $h_{TD}$, using its MAC key $k$. These hashes, $\mathcal{O}$ and the MAC are passed via \textit{Out()}, reflecting transmission via the VMM to QE. A~\textit{ReportReady} fact is also produced to maintain the internal binding between the operation's values and the TD's secret key.

For \quote generation (Rule~\ref{tamarin:rule:quote_gen_inmem_public}), QE receives \tdreport via \textit{In()} and recomputes the MAC to verify it before signing. 
Upon successful verification, QE signs the hashes of the $\mathcal{I}$, $op$, $\chal$, $\mathcal{O}$, and $h_{TD}$ using $qsk$, and send \textit{Out()} the signed \quote and $\mathcal{O}$.

Lastly, for Rule~\ref{tamarin:rule:quote_ver_inmem_public}, \vrf consumes \quote from the network alongside the known QE public key, checking the signature against the expected hash values. The~\textit{VrfAccepts} action fact records the outcome, binding \vrf's acceptance to the specific $\mathcal{I}$, $op$, $\chal$, $\mathcal{O}$, $h_{TD}$, and public key.

\subsection{Security Proofs for Case 1}\label{appdx:tamarin-case1-lemmas}


\begin{figure}[t]
\begin{lstlisting}[style=tamarinstyle]
// Case 1: In-memory and public outputs
|\lemmaheader{Key Secrecy}\label{tamarin:lemma:key_sec}|
lemma key_secrecy [reuse]:
  "All k #i. SecretKey(k) @ #i ==> not (Ex #j. K(k) @ #j)"

|\lemmaheader{TD Measurement Integrity}\label{tamarin:lemma:td_integrity}|
lemma td_measurement_integrity:
  "All op I chal O tdImage vk #n.
      VrfAccepts(op, I, chal, O, tdImage, true, vk) @ #n
    ==> Ex #i. InitialInstall(tdImage) @ #i & #i < #n"

|\lemmaheader{Output came from Operation and Input}\label{tamarin:lemma:o_from_op}|
lemma O_came_from_op_and_I:
  "All op I chal O hTD pkQ #i.
     VrfAccepts(op, I, chal, O, hTD, true, pkQ) @ #i
     ==> O = runOp(op, I)"

|\lemmaheader{Quote Authenticity}\label{tamarin:lemma:quote_authenticity}|
lemma quote_authenticity [reuse]:
  "All op I chal O hTD vk #n.
     VrfAccepts(op, I, chal, O, hTD, true, vk) @ #n
   ==> Ex #m. QEQuotes(h(I), h(op), h(chal), h(O), hTD, vk) @ #m & #m < #n"

|\lemmaheader{Quote Only If MAC is correct}\label{tamarin:lemma:quote_gen_if_mac}|
lemma quote_only_if_mac:
  "All hI hOp hChal hO hTD mac k #i.
      MacCheck(hI, hOp, hChal, hO, hTD, mac, k) @ #i
    ==> mac = h(<hI, hOp, hChal, hO, hTD, k>)"

|\lemmaheader{Input Measurement Ingetrity}\label{tamarin:lemma:input_integrity}|
lemma input_measurement_integrity:
  "All op I chal O hTD vk #n.
      VrfAccepts(op, I, chal, O, hTD, true, vk) @ #n
    ==> Ex #j. InputMessage(op, I, chal) @ #j & #j < #n"
\end{lstlisting}
\end{figure}

Lemmas~\ref{tamarin:lemma:key_sec}--~\ref{tamarin:lemma:input_integrity} are used to prove the protocol model detailed in Appendix~\ref{appdx:tamarin-case1} uphold security goals \textbf{[G1-6]} from Appendix~\ref{appdx:tamarin_sec_goal}: 
\begin{enumerate}[leftmargin=*,label=\textbf{[G\arabic*]},nosep]
    \item \textbf{Confidentiality of Secrets}: Lemma~\ref{tamarin:lemma:key_sec} ensures that any key marked as secret at initialization is never derivable by \adv;
    \item \textbf{TD Measurement Integrity:} Lemma~\ref{tamarin:lemma:td_integrity} ensures that the \vrf only accepts an attestation if the TD image referenced in the attestation was legitimately installed prior to verification, binding accepted attestations to a known, measured software state.
    Notably, the integrity guarantee is scoped to the TD image measurement. Whether the expected TD image was used requires consulting reference values from \tauth, as described in Section~\ref{sec:attestation}. Note that due to SPDM occurring to establish a secure channel with GPU at the time of TD creation~\cite{spdm,gu2025nvidia}, these takeaways extend to the TD+GPU environment;
    \item \textbf{Quote Authenticity:} Lemma~\ref{tamarin:lemma:quote_authenticity} ensures that \vrf only accepts a \quote originating from \prv's QE, ruling out any scenario where a valid attestation is produced without QE's involvement;
    \item \textbf{Quote Integrity:} Lemma~\ref{tamarin:lemma:quote_gen_if_mac} ensures that QE invokes quote generation only when the received MAC is well-formed, binding quote generation to a MAC computed correctly over the expected hash values and the TD's secret key, and guaranteeing that \tdreport has not been tampered. Together, these two lemmas establish an unbroken chain of trust from the TD's internal state through QE to \vrf;
    \item \textbf{Output Integrity}: Lemma~\ref{tamarin:lemma:o_from_op} ensures that whenever \vrf accepts an attestation, the reported output $\mathcal{O}$ is exactly the result of applying the operation op to input $I$ via \textit{runOp(op, I)} in the TD.
    This rules out any scenario in which verification passes with a fabricated or mismatched result produced outside of the TD;
    \item \textbf{Input Measurement Integrity:} Lemma~\ref{tamarin:lemma:input_integrity} ensures that for every \textit{VrfAccepts} event, there must exist an \textit{InputMessage} event carrying the same $op$, $I$, and $chal$. Since \textit{InputMessage} is only emitted by \textit{Get\_Operation\_Inputs}, which runs inside the TD, and the subsequent \textit{Execute\_Operation} and \textit{TD\_Produce\_Report} rules consume and propagate those exact values through internal linear facts before any output is produced, the accepted quote is guaranteed to have been computed over inputs that were loaded into TD-protected memory.
\end{enumerate}











\subsection{Case 2: In-memory Private Outputs}\label{appdx:tamarin-case-2}

Here we describe the changes to the rules/lemmas for Case 1:~\ref{appdx:tamarin-case1}--\ref{appdx:tamarin-case1-lemmas} to enable Case 2 that accounts for private outputs.

\begin{figure}[t]
\begin{lstlisting}[style=tamarinstyle]
// Case 2: In-memory and private outputs
|\ruleheader{Operation Execution (memory-mapped)}\label{tamarin:rule:op_exec_inmem_private}|
rule Execute_Operation:
  let
    O = runOp(op, I)
  in
  [ OperationInputs(op, I, chal, hTD)]
  --[ RunOperation(op, I, chal, O), |\hl{SecretOutput(O)}| ]
  [ CompleteOperation(op, I, chal, O, hTD) ]

|\ruleheader{TD Report Generation}\label{tamarin:rule:tdreport_gen_inmem_private}|
rule TD_Produce_Report:
    let
      O = runOp(op, I)
      mac = h(<h(I), h(op), h(chal), h(O), hTD, k>)
    in
    [!Key(k), CompleteOperation(op, I, chal, runOp(op, I), hTD)]
  --[TDProducesReport(op, I, chal, runOp(op, I), hTD, k)]->
    [Out(<|\hl{h(I)}|, h(op), h(chal), h(O), hTD, mac>), 
      ReportReady(op, I, chal, runOp(op, I), hTD, k)]

|\ruleheader{Quote Generation}\label{tamarin:rule:quote_gen_inmem_private}|
rule QE_Produce_Quote:
    let 
      mac = h(<hI, hOp, hChal, hO, hTD, k>)
      |\hl{hashes = <hI, hOp, hChal, hO, hTD>}|
      |\hl{sig = sign(hashes, qsk)}|
    in
    [ !QESignKey(qsk), In(<hI, hOp, hChal, hO, hTD, mac>), ReportReady(op, I, chal, O, hTD, k)
    ]
  --[ MacCheck(hI, hOp, hChal, hO, hTD, mac, k), QEQuotes(hI, hOp, hChal, hO, hTD, pk(qsk)) ]->
    [Out(<|\hl{hashes}|, sig>), SignatureReady(op, I, chal, O, hTD, sig)]

|\ruleheader{Verification}\label{tamarin:rule:quote_ver_inmem_private}|
rule Vrf_Verifies:
    let 
      |\hl{hashes = <hI, hOp, hChal, hO, hTD>}|
    in
    [ !QEVerKey(pkQ),
      SignatureReady(op, I, chal, O, hTD, sig),
      In(<|\hl{hashes}|, sig>) ]
  --[ VrfAccepts(op, I, chal, O, hTD, verify(sig, <h(I), h(op), h(chal), h(O), hTD>, pkQ), pkQ) ]->
    [] 
\end{lstlisting}
\end{figure}

\textbf{Modifications to Rules.} The key change from public to private output is how $\mathcal{O}$ is exposed. In Case 1, \textit{TD\_Produce\_Report} outputs $\mathcal{O}$ directly to the network. In Case 2, it outputs only the hash of it, keeping $\mathcal{O}$ hidden. \textit{Execute\_Operation} also adds \textit{SecretOutput(O)} action fact to mark $\mathcal{O}$ for privacy tracking. 

\textbf{Modifications to Lemmas.} Goals \textbf{[G2-6]} in Case 2 are upheld using the same lemmas from Case 1, with one additional lemma added to ensure that~\textbf{[G1]} still holds. Lemma~\ref{tamarin:lemma:private_o_inmem} ensures that $\mathcal{O}$ remains secret. It states that any value marked \textit{SecretOutput} is never learned by \adv, which makes it satisfy \textbf{[G1]}.

\begin{figure}[t]
\begin{lstlisting}[style=tamarinstyle]
// Case 2: In-memory and private outputs
|\lemmaheader{Private Output}\label{tamarin:lemma:private_o_inmem}|
lemma private_O [reuse]:
  "All X #i. SecretOutput(X) @ #i ==> not (Ex #j. K(X) @ #j)" 
\end{lstlisting}
\end{figure}

\subsection{Case 3: Memory-mapped Public Outputs}\label{appdx:tamarin-case-3}

Here we describe only the changes to Case 1 protocol from~\ref{appdx:tamarin-case1}--\ref{appdx:tamarin-case1-lemmas} to enable Case 3 protocol by accounting for memory-mapped datasets with public outputs. The changes to the model are highlighted in yellow in Rules~\ref{tamarin:rule:exec-mem-mapped-1}--\ref{tamarin:rule:use-disk-item} and Lemmas~\ref{tamarin:lemma:input_integrity_mm}--\ref{tamarin:lemma:ms_O_match}.


\begin{figure}[t]
\begin{lstlisting}[style=tamarinstyle]
// Case 3: Mem-mapped and public outputs
|\ruleheader{Execute Operation (mem-mapped)}\label{tamarin:rule:exec-mem-mapped-1}|
rule Execute_Operation:
  [ OperationInputs(op, chal, hTD), |\hl{Fr(\textasciitilde{}sid)}| ]
  --[ |\hl{SessionStarted}|(~sid, op, chal, hTD) ]->
  [ |\hl{State}|(~sid, op, chal, hTD, runOp(op, 'empty'), 'empty', 'none', %1)} ]

|\ruleheader{Execution step (lookup from disk)}\label{tamarin:rule:exec-disk-lookup}|
|\hl{rule Execution\_lock\_and\_lookup}|:
  [ State(sid, op, chal, hTD, pretrained, block_data_multiset, block_indices, %index) ]
  -->
  [ Out(<'get', sid, %index>), PendingLookup(sid, op, chal, hTD, pretrained, block_data_multiset, block_indices, %index) ]

|\ruleheader{Return item from disk}\label{tamarin:rule:get-from-disk}|
|\hl{rule DatasetStorage}|:
  [ In(<'get', sid, index>) ]
  -->
  [ Out(<'batch', sid, datasets(index)> ) ]

|\ruleheader{Execution step (use disk item)}\label{tamarin:rule:use-disk-item}|
|\hl{rule Execution\_unlock\_and\_continue}|:
  let
    model = runOp(op, batch)
    new_block_data_multiset = block_data_multiset ++ <%index, batch>
    new_block_indices = h(<batch, block_indices>)
  in
  [ In(<'batch', sid, batch>), PendingLookup(sid, op, chal, hTD, pretrained, block_data_multiset, block_indices, %index) ]
  --[ Trained(sid, op, chal, hTD,  model, new_block_data_multiset, new_block_indices) ]->
  [ State(sid, op, chal, hTD, model, new_block_data_multiset, new_block_indices, %index %+ %1) ]


\end{lstlisting}
\end{figure}

\textbf{Modifications to Rules.} From Case 1, the rules for TD Initialization (Rule~\ref{tamarin:rule:init_td}) and Get Inputs (Rule~\ref{tamarin:rule:get_inputs_inmem_public}) are also in the model for Case 3. However, Case 3 requires the remaining rules to change, plus the addition of three new rules. 

Many changes to the operation execution are to model an incremental operation that has several steps: first, starting the execution, making a request to disk, modeling the disk itself, and finalizing that step (and the multiset). 
Rule~\ref{tamarin:rule:exec-mem-mapped-1} shows the first part of this process, in which a session ID is specified (\textit{\textasciitilde{}sid}) along with an action fact \textit{SessionStarted} to log this event. The postcondition of this rule is a fact that specifies the initial state. Notably, the \textit{State} fact contains a field for output, and in this initial state, it is equal to performing the \textit{runOp()} on an empty dataset. 

Rule~\ref{tamarin:rule:exec-disk-lookup} specifics the first step for execution with a memory-mapped dataset. A record of the dataset is requested from the disk by specifying the index. This rule results in sending \textit{Out} a get request from disk for the particular index, along with a fact to signify a pending lookup (\textit{PendingLookup}).

Rule~\ref{tamarin:rule:get-from-disk} specifies a simple rule to simulate fetching an item from the disk. The get request is received from \textit{In}, and a batch that was fetched from the index is returned through \textit{Out}. Both messages from this rule pass through an untrusted channel to simulate \adv control over the disk. 

Rule~\ref{tamarin:rule:use-disk-item} specifies the continuation of the execution that uses the fetched batch from the requested index, and the accumulation of the multiset that is used to construct the \MSH. The multiset construction is shown by accumulating \textit{$<\%index, batch>$} into the multiset and hashing it. This also creates a postcondition containing fact \textit{State}, resulting in the possibility of re-evaluation of Rule~\ref{tamarin:rule:exec-disk-lookup} (e.g., to use more data records and grow the multiset further) or to continue to report generation.

The rules that specify \tdreport Generation, Quote Generation, and Verification are largely the unmodified Rule~\ref{tamarin:rule:tdreport_gen_inmem_public},~\ref{tamarin:rule:quote_gen_inmem_public}, and~\ref{tamarin:rule:quote_ver_inmem_public}, respectively. A very minor change is that the \MSH measurement is included along with all other hashes in each fact and action fact. 

\begin{figure}[t]
\begin{lstlisting}[style=tamarinstyle]
// Case 3: Mem-mapped and public outputs
|\lemmaheader{Input Measurement Integrity}\label{tamarin:lemma:input_integrity_mm}|
lemma input_measurement_integrity:
"All op I hI chal O hTD vk #n.
  VrfAccepts(op, I, hI, chal, O, hTD, true, vk) @ #n
  & not( I = 'empty')
 ==> Ex indices sid #j.
 Trained(sid, op, chal, hTD, O, I, indices) @ #j"

|\lemmaheader{Equal sampling sequences implies equal multiset}\label{tamarin:lemma:msh-implies-multiset}|
lemma Seq_Multiset_Match [use_induction,reuse]: all-traces
"All sid1 sid2 m1 m2 hm1 hm2 hi op1 op2 chal1 chal2 hTD1 hTD2 #i #j.
 Trained(sid1, op1, chal1, hTD1, m1, hm1, hi) @ i 
& Trained(sid2, op2, chal2, hTD2, m2, hm2, hi) @ j
==> hm1 = hm2"

|\lemmaheader{Equal multiset implies equal input sample sequence}\label{tamarin:lemma:ms_hash_match}|
restriction MS_Seq_Match:
"All sid1 sid2 op1 op2 chal1 chal2 hTD1 hTD2 m1 m2 hm hi1 hi2 #i #j. 
Trained(sid1, op1, chal1, hTD1, m1, hm, hi1) @ i 
& Trained(sid2, op2, chal2, hTD2, m2, hm, hi2) @ j
    ==> hi1 = hi2"

|\lemmaheader{Equal input sample sequence implies equal output}\label{tamarin:lemma:msh_o_match}|
lemma Seq_O_Match [use_induction,reuse]: all-traces
"All sid1 sid2 op chal1 chal2 hTD1 hTD2 m1 m2 hm1 hm2 hi #i #j. 
Trained(sid1, op, chal1, hTD1, m1, hm1, hi) @ i 
& Trained(sid2, op, chal2, hTD2, m2, hm2, hi) @ j 
==> m1 = m2"

|\lemmaheader{Equal multiset implies equal output}\label{tamarin:lemma:ms_O_match}|
lemma MS_O_match [reuse]: all-traces
"All sid1 sid2 op chal1 chal2 hTD1 hTD2 m1 m2 hm hi1 hi2 #i #j. 
 Trained(sid1, op, chal1, hTD1, m1, hm, hi1) @ i 
 & Trained(sid2, op, chal2, hTD2, m2, hm, hi2) @ j 
 ==> m1 = m2"
\end{lstlisting}
\end{figure}

\textbf{Modifications to Lemmas.} One lemma from Case 1 is modified, and three more are introduced to account for the multiset construction.
\textbf{[G1-4]} use Lemmas~\ref{tamarin:lemma:key_sec}, ~\ref{tamarin:lemma:td_integrity}, ~\ref{tamarin:lemma:quote_authenticity}, and~\ref{tamarin:lemma:quote_gen_if_mac}, respectively, as discussed in Appendix~\ref{appdx:tamarin-case1-lemmas}. \textbf{[G5-6]} are upheld by the the following lemmas:
\begin{enumerate}[leftmargin=*,label=\textbf{[G\arabic*]},nosep]
    \item \textbf{Output Integrity}: uses Lemma~\ref{tamarin:lemma:o_from_op}, discussed in Appendix~\ref{appdx:tamarin-case1-lemmas} along with new Lemmas~\ref{tamarin:lemma:msh_o_match} and~\ref{tamarin:lemma:ms_O_match}. 
    Lemma~\ref{tamarin:lemma:msh_o_match} provides that two training operations that produce equivalent sampling sequences (denoted \textit{hI}) result in the same output (e.g., \textit{m1} and \textit{m2}). Similarly, Lemma ~\ref{tamarin:lemma:ms_O_match} establishes that two training operations that use the same input multiset (\textit{hm}) results in the same output;
    \item \textbf{Input Integrity}: uses Lemmas~\ref{tamarin:lemma:input_integrity_mm},~\ref{tamarin:lemma:msh-implies-multiset}, and ~\ref{tamarin:lemma:ms_hash_match}. Lemma~\ref{tamarin:lemma:input_integrity_mm} is a modified version of Lemma~\ref{tamarin:lemma:input_integrity} that checks for the \textit{Trained} event rather than \textit{InputMessage}. In addition, since the dataset is measured incrementally, the integrity check does not apply when the input is empty. Given these two modifications, this lemma provides that a \quote passing verification implies that the input dataset is not empty and that it was indeed used for training. 
    Lemma~\ref{tamarin:lemma:msh-implies-multiset} provides that given two training operations have occurred with equal sampling sequences, the two inputs must be equal. Finally, Lemma~\ref{tamarin:lemma:ms_hash_match} specifies that two training operations with the same recorded multiset must have used the same sampling sequences. This lemma uses the keyword \textit{restriction}, meaning that it must be proven by hand. The proof for this is shown in Appendix~\ref{appdx:msh-security-reduction}.
\end{enumerate}

\subsection{Case 4: Memory-mapped Private Outputs}\label{appdx:tamarin-case-4}

Here we describe the changes to the Case 3 model from Appendix~\ref{appdx:tamarin-case-3} to account for Case 4 with private outputs and memory-mapped datasets. 

\textbf{Modifications to Rules.} Only one rule requires modification, and that is Rule~\ref{tamarin:rule:report-gen-mm-private} for \tdreport Generation. Here, secret output $\mathcal{O}$ is logged using the action fact \textit{SecretOutput} along with $op$ and $\mathcal{I}$ that were used to produce $\mathcal{O}$. Additionally, the multiset hash (\textit{hI}) is included in the set of output hash values sent via \textit{Out}.

\textbf{Modifications to Lemmas.} Goals \textbf{[G2-6]} are upheld using the same lemmas as Case 3 protocol, described in Appendix~\ref{appdx:tamarin-case-3}. Goal \textbf{[G1]} is supported by the verification of Lemma~\ref{tamarin:lemma:key_sec} for key secrecy, as discussed in Appendix~\ref{appdx:tamarin-case1-lemmas}, and it is also supported by the verification of Lemma~\ref{tamarin:lemma:private_o_mem_mapped}, which specifies that $\mathcal{O}$ cannot be learned given any operation and non-empty input dataset. 

\begin{figure}[t]
\begin{lstlisting}[style=tamarinstyle]
// Case 4: Mem-mapped and private outputs
|\ruleheader{TD Report Generation (mem-mapped)}\label{tamarin:rule:report-gen-mm-private}|
rule TD_Produce_Report:
    let
    O = runOp(op, I)
    mac = h(<hI, h(op), h(chal), h(O), hTD, k>)
    in
    [ !Key(k), State(sid, op, chal, hTD, O, I, hI, index) ]
  --[ TDProducesReport(op, I, hI, chal, O, hTD, k), 
      |\hl{SecretOutput(O, op, I)}| 
    ]->
    [ Out(<|\hl{hI}|, h(op), h(chal), h(O), hTD, mac>), 
      ReportReady(op, I, hI, chal, O, hTD, k) ]



\end{lstlisting}
\end{figure}

\begin{figure}[t]
\begin{lstlisting}[style=tamarinstyle]
// Case 4: Mem-mapped and private outputs
|\lemmaheader{Private Output}\label{tamarin:lemma:private_o_mem_mapped}|
lemma private_O:
  "All O I op #i. SecretOutput(O, op, I) @ #i 
  ==> not ( Ex #j. not (
      O = train(op, 'empty')
    ) & K(O) @ #j )"
\end{lstlisting}
\end{figure}

\begin{table}[t]
    \centering
    \caption{Statistics for lemma verification: Max Steps denotes the highest steps for any lemma; Total Steps is the sum of steps across all lemmas; Peak Mem. is the maximum RAM usage; Time is the total verification run-time.}
    \vspace{-1em}
    \resizebox{\columnwidth}{!}{
    \begin{tabular}{l!{\vruletab}c!{\vruletab}c!{\vruletab}c!{\vruletab}c}
    \bottomrule
    
    \toprule
         \textbf{Protocol} & \textbf{Max Steps} & \textbf{Total Steps} & \textbf{Peak Mem. (MiB)} & \textbf{Time (s)} \\
    \midrule
         Case 1 & 20 & 38 & 26799 & 89.2\\
         Case 2 & 16 & 38 & 246 & 2.33 \\
         Case 3 & 42 & 98 & 2268 & 65.7\\
         Case 4 & 42 & 115 & 270 & 5.10\\
    \bottomrule
    
    \toprule
    \end{tabular}
    }
    \label{tab:verification}
\end{table}

\subsection{Verification Results}

Each protocol case has the lemmas outlined in Appendices~\ref{appdx:tamarin-case1-lemmas}--~\ref{appdx:tamarin-case-4}, along with several ``helper lemmas'' that test functionality of the protocol.
Table~\ref{tab:verification} shows performance statistics when using Tamarin to analyze the four protocol cases and prove all lemmas (functional and security). Case 2 is the most simplistic for two reasons. First, it uses a simpler model of the in-memory dataset. Second, it has fewer symbols that are controllable by \adv, resulting in less exploration by Tamarin. Both Case 3 and Case 4 require verifying the lemma that takes the most steps for verification: Lemma~\ref{tamarin:lemma:msh-implies-multiset}, which requires 42 steps.
The higher memory usage and run-time of Case 1 are due to a functional helper lemma that forces the prover to evaluate each rule. 
In this case, it must simultaneously solve for execution semantics of the custom function that is used to produce $\mathcal{O}$ plus the cryptographic abstraction \textit{hO}. In Case 2, execution semantics do not require as in-depth an exploration since \adv only learns \textit{hO}.


\subsection{Multiset hashing security reduction}\label{appdx:msh-security-reduction}
\method's ability to access datasets out-of-order that cannot fit all at once into the TEE's memory depends on its use of an incremental multiset hash \MSH satisfying the homomorphism
\[\MSH(M_1 \cup M_2) = \MSH(M_1) \cdot \MSH(M_2).\]

Specifically, given an ordered dataset $\mathcal{D} = (d_1, d_2, \ldots, d_n)$ of $n$ records, we require that the TEE, after receiving records $p_1, \ldots, p_n$, can be assured that the values $d_1', \ldots, d_n'$ that it  receives are the same values $d_i' = d_{p_i}$ in the original dataset, so long as
\[
\MSH\left(\left\{ (1, d_1), \ldots, (n, d_n) \right\}\right) = \MSH\left(\left\{ (p_1, d_1'), \ldots, (p_n, d_n') \right\}\right) .
\]
The left-hand side can be easily precomputed whenever $p$ is a permutation; for convenience, we denote this hash $\MSH(\mathcal{D})$.

The adversary's goal is defined in the following game in which the adversary \aadv\ attempts to provide the TEE with records not matching $\mathcal{D}$, shown in \Cref{fig:games}. Our goal is to reduce this to the \MSH collision game $G_\textrm{\MSH-Collision}^{\badv}$, shown in \Cref{fig:msh-collision-game}.

\begin{figure}
\begin{tikzpicture}
\node[draw, rounded corners, align=left, text width=0.95\linewidth] {
\textbf{$G_\textrm{\MSH-Collision}^{\badv_\aadv}$}\\
\vspace{0.25em}
\hrule width \linewidth
\vspace{0.25em}
$M_1, M_2 \xleftarrow{\$} \badv_\aadv(n)$\\
\textbf{if} $\MSH(M_1) = \MSH(M_2) \wedge M_1 \ne M_2$\\
\quad \textbf{return} 1\\
\textbf{else}\\
\quad \textbf{return} 0\\
\textbf{endif}
};
\end{tikzpicture}
\vspace{-1em}
\caption{\MSH collision game.}\label{fig:msh-collision-game}
\end{figure}

\begin{figure}
\centering

\begin{tikzpicture}
\node[draw, rounded corners, align=left, text width=0.95\linewidth] {
\textbf{$G_\textrm{Streamed-Incorrect-Data}^{\mathcal{D},\aadv}$}\\
\vspace{0.25em}
\hrule width \linewidth
\vspace{0.25em}
H $\leftarrow$ \MSH($\{\}$)\\
\textbf{for} $i = 1 \ldots n$\\
\quad $(p_i, d_i') \xleftarrow{\$} \aadv()$\\
\quad $H \leftarrow H \cdot \MSH(\{(p_i, d_i')\})$\\
\textbf{endfor}\\
\textbf{if} $H = \MSH(\mathcal{D}) \wedge \exists i : d_i' \neq d_{p_i}$\\
\quad \textbf{return} 1\\
\textbf{else}\\
\quad \textbf{return} 0\\
\textbf{endif}
};
\end{tikzpicture}

\centering $\Downarrow$ {\small Reformulated using \MSH homomorphism}
\vspace{0.25em}

\begin{tikzpicture}
\node[draw, rounded corners, align=left, text width=0.95\linewidth] {
\textbf{$G_{\textrm{Streamed-Incorrect-Data}'}^{\mathcal{D},\aadv}$}\\
\vspace{0.25em}
\hrule width \linewidth
\vspace{0.25em}
$M_1, M_2 \xleftarrow{\$} \badv_\aadv(n):$\\
\hspace{4em} \textbf{for} $i = 1 \ldots n$\\
\hspace{4em} \quad $p_i, d_i' \xleftarrow{\$} \aadv()$\\
\hspace{4em} \textbf{endfor}\\
\hspace{4.2em}\textbf{return} $\{(1, d_1), \ldots, (n, d_n) \}, \{ (p_1, d_1'), \ldots, (p_n, d_n') \}$ \\
\textbf{if} $\MSH(M_1) = \MSH(M_2) \wedge \exists i : d_i' \neq d_{p_i}$\\
\quad \textbf{return} 1\\
\textbf{else}\\
\quad \textbf{return} 0\\
\textbf{endif}
};
\node[draw, dashed, fill=none, align=left, text width=0.77\linewidth, minimum height=2.0cm] at (0.64,0.5) {};
\draw (-2.65, 1.14) -- (3.90, 1.140);
\end{tikzpicture}

\centering $\Downarrow$
{\small $\badv_\aadv$ also wins}
\vspace{0.25em}

\begin{tikzpicture}
\node[draw, rounded corners, align=left, text width=0.95\linewidth] {
\textbf{$G_\textrm{\MSH-Collision}^{\badv_\aadv}$}\\
\vspace{0.25em}
\hrule width \linewidth
\vspace{0.25em}
$M_1, M_2 \xleftarrow{\$} \badv_\aadv(n)$\\
\textbf{if} $\MSH(M_1) = \MSH(M_2) \wedge M_1 \ne M_2$\\
\quad \textbf{return} 1\\
\textbf{else}\\
\quad \textbf{return} 0\\
\textbf{endif}
};
\end{tikzpicture}
\vspace{-1em}
\caption{Game hops to prove Theorem~D.1.}
\label{fig:games}
\end{figure}

\begin{theorem}[Incremental hashing of streamed data.]\label{thm:hashing}
For any adversary \aadv and dataset $\mathcal{D} = (d_1, d_2, \ldots, d_n)$, there exists $\badv_\aadv$ such that:
\begin{align*}
    \Pr{(G_\textrm{Streamed-Incorrect-Data}^{\mathcal{D},\aadv})}{} \le \Pr{(G_\textrm{\MSH-Collision}^{\badv_\aadv})}{}
\end{align*} 
\end{theorem}
\begin{proof}
We observe that $G_\textrm{Streamed-Incorrect-Data}^{\mathcal{D},\aadv}$ can be restructured in terms of a new adversary $\badv_\aadv$, shown in \Cref{fig:games}.    This game is very similar to $G_\textrm{\MSH-Collision}^{\badv_\aadv}$, except that the requirement $M_1 \ne M_2$ is replaced by $\exists i : d_i' \neq d_{p_i}$.  That is, there exists some $i$ such that there exist distinct elements $(p_i, d_{p_i}) \in M_1$ and $(p_i, d_i') \in M_2$.
Since $M_1$ is defined as
\[
M_1 = \{ (1, d_1), \ldots, (n, d_n) \},
\]
it contains exactly one element having any particular index $i \in \{1, \ldots, n\}$, and none with indices outside this range.   Since $(p_i, d_i') \ne (p_i, d_{p_i})$, $M_1$ and $M_2$ therefore differ at the element with this index, and so $\badv_\aadv$ also wins $G_\textrm{\MSH-Collision}^{\badv_\aadv}$. 
\end{proof}

\end{document}